\documentclass[aps,prx,twocolumn,notitlepage,showpacs,amsmath,amstex,amssymb,citeautoscript,longbibliography]{revtex4-1}
\pdfoutput=1
\usepackage{natbib}
\usepackage[english]{babel}
\usepackage{letltxmacro}
\usepackage{latexsym}
\LetLtxMacro{\ORIGselectlanguage}{\selectlanguage}
\makeatletter
\DeclareRobustCommand{\selectlanguage}[1]{%
  \@ifundefined{alias@\string#1}
    {\ORIGselectlanguage{#1}}
    {\begingroup\edef\x{\endgroup
       \noexpand\ORIGselectlanguage{\@nameuse{alias@#1}}}\x}%
}
\newcommand{\definelanguagealias}[2]{%
  \@namedef{alias@#1}{#2}%
}
\makeatother
\definelanguagealias{en}{english}
\definelanguagealias{English}{english}
\usepackage{graphicx}
\usepackage{amsmath}
\usepackage{amsfonts}
\usepackage{amssymb,bbding}
\usepackage{bm}
\usepackage{color}
\usepackage[percent]{overpic}
\usepackage{soul} 
\usepackage{amssymb}
\usepackage{wasysym}
\usepackage{dsfont}
\usepackage{float}
\usepackage{braket}
\usepackage{cancel}
\usepackage{comment}
\usepackage{hyperref}
\hypersetup{
    bookmarks=false,         
    unicode=false,          
    pdftoolbar=false,        
    pdfmenubar=true,        
    pdffitwindow=false,     
    pdfstartview={FitH},    
    pdftitle={Slow quantum thermalization and many-body revivals from classical mixed phase space},    
    pdfauthor={A. A. Michailidis, C. J. Turner, Z. Papic, D. A. Abanin, and M. Serbyn},     
    pdfsubject={},   
    pdfcreator={},   
    pdfproducer={}, 
    pdfkeywords={quantum scars} {non-ergodic dynamics}{mixed phase space}{quantum many-body chaos}{thermalization}, 
    pdfnewwindow=true,      
    colorlinks=true,       
    linkcolor=black,          
    citecolor=blue,        
    filecolor=magenta,      
    urlcolor=blue           
}

\newcommand{\be}{\begin{equation}}
\newcommand{\ee}{\end{equation}}
\newcommand{\bea}{\begin{eqnarray}}
\newcommand{\eea}{\end{eqnarray}}

\renewcommand{\vec}[1]{{\bf #1}}

\newcommand{\corr}[1]{\langle{ #1}\rangle}

\newcommand{\SOM}{\cite{SOM}}
\begin{document}
\title{Slow quantum thermalization and many-body revivals from mixed phase space} 
\author{A. A. Michailidis$^1$, C. J. Turner$^2$, Z. Papi\'c$^2$, D. A. Abanin$^3$, and M. Serbyn$^1$}
\affiliation{$^1$IST Austria, Am Campus 1, 3400 Klosterneuburg, Austria}
\affiliation{$^2$School of Physics and Astronomy, University of Leeds, Leeds LS2 9JT, United Kingdom}
\affiliation{$^3$Department of Theoretical Physics, University of Geneva,
24 quai Ernest-Ansermet, 1211 Geneva, Switzerland}
\date{\today}
\begin{abstract}
Relaxation of few-body quantum systems can strongly depend on the initial state when the system's semiclassical phase space is  \emph{mixed}, i.e., regions of chaotic motion coexist with regular islands. In recent years, there has been much effort to understand the process of thermalization in strongly interacting quantum systems that often lack an obvious semiclassical limit.
Time-dependent variational principle (TDVP) allows to systematically derive an effective classical (nonlinear) dynamical system by projecting unitary many-body dynamics onto a manifold of weakly-entangled variational states.  We demonstrate that such dynamical systems generally possess mixed phase space. When TDVP errors are small, the mixed phase space leaves a footprint on the exact dynamics of the quantum model. For example, when the system is initialized in a state belonging to a stable periodic orbit or the surrounding regular region, it exhibits persistent many-body quantum revivals. As a proof of principle, we identify new types of ``quantum many-body scars", i.e., initial states that lead to long-time oscillations in a model of interacting Rydberg atoms in one and two dimensions.  Intriguingly, the initial states that give rise to most robust revivals are typically entangled states.
On the other hand, even when TDVP errors are large, as in the thermalizing tilted-field Ising model, initializing the system in a regular region of phase space leads to surprising slowdown of thermalization. Our work establishes TDVP as a method for identifying interacting quantum systems with anomalous dynamics in arbitrary dimensions. Moreover, the mixed-phase space classical variational equations allow to find slowly-thermalizing initial conditions in interacting models.  Our results shed light on a link between classical and quantum chaos, pointing towards possible extensions of classical Kolmogorov-Arnold-Moser (KAM) theorem to quantum systems.
\end{abstract}
\maketitle

\section{Introduction}

Technological advances in synthetic quantum systems~\cite{PolkovnikovRMP,BlochColdAtoms} have started an era where  non-equilibrium dynamics of isolated quantum matter can be experimentally probed. The process of intrinsic thermalization that may occur in isolated many-body systems results in featureless thermal states and  scrambling of quantum information. Systems which avoid thermalization due to an extensive number of integrals of motion, such as many-body localized (MBL)~\cite{AbaninRev} and Bethe-ansatz integrable systems~\cite{Sutherland}, may host non-trivial quantum-coherent states, which have been fruitfully studied in recent years. However, these systems require either the presence of quenched disorder (MBL) or fine tuning (Bethe-ansatz integrability). Therefore, it is desirable to find other routes towards extending quantum coherence in many-body systems.  

Progress towards extending quantum coherence is tied with developing a more complete understanding of quantum many-body chaos and thermalization~\cite{Polkovnikov-rev}. In this direction, recent studies~\cite{Nahum18,Chalker18,Prosen18} considered a model of ``typical'' quantum dynamics generated by applying random unitary operators in a quantum circuit, which allowed to obtain results for the dynamics of entanglement and other physical observables. In such models, thermalization rate should not depend strongly on the initial state. Recent experiments, however, revealed that the dynamics in physical many-body systems may be significantly more complex~\cite{Bloch02,Bernien2017}. In particular, for certain initial states, the chain of interacting Rydberg atoms~\cite{Bernien2017} was found to exhibit much slower thermalization or even its absence on the experimentally accessible time scales. Dynamics following from such initial states features persistent periodic revivals of local observables, such as the density of domain walls in the initial state as a function of time. 

The observed periodic revivals were explained by ``quantum many-body scars''~\cite{Turner2017,Turner2018,wenwei18} --  a subset of anomalous  eigenstates with strongly non-thermal properties~\cite{Turner2017}. Many-body scarring is a generalization of the well-known phenomenon of quantum scars in stadium billiards, where anomalous eigenstates exhibit an increased concentration of probability density around unstable classical periodic orbits~\cite{Heller84}. Atypical eigenstates had previously been constructed analytically in the AKLT spin chain~\cite{Bernevig2017,BernevigEnt}, and it has been suggested that some of these non-thermalizing states may be closely related to the ones in the Rydberg atom chain~\cite{Motrunich17, Shiraishi2019}. Additionally, various types of non-thermalizing behaviors have since been reported in a number of other models~\cite{Calabrese16,Motrunich17,Konik_Robinson2018,Mori17,*Rigol18,Titus19,Pretko19,Nandkishore19,Pollmann19,Papic19}.

In the context of \emph{few-body} quantum chaos~\cite{HaakeBook}, the strong dependence of quantum relaxation rate on the initial state more often emerges not from unstable periodic orbits (as in quantum scars), but from the phenomenon of \emph{mixed phase space}~\cite{Percival,Bohigas}.  According to Kolmogorov-Arnold-Moser (KAM) theorem~\cite{Kolmogorov,*Arnold}, weak deformations of an integrable classical system destroy the regular structure of its phase space only in some regions, leading to a coexistence of regular islands with regions of chaotic motion~\cite{Zaslavsky}. The semiclassical limit allows to introduce a notion of regular and chaotic eigenstates that are dominated by the corresponding regions of phase space~\cite{Percival,Bohigas}. The regular eigenstates strongly affect the dynamics. When a quantum system is initialized in a wave packet residing predominantly in the mixed region of the phase space, it exhibits much slower relaxation compared to the case when the quantum evolution begins in the chaotic region of the phase space~\cite{HaakeBook}. 

In contrast to few-body quantum systems, in the many-body case one expects that conditions for KAM theorem become very stringent, potentially leading to a quick disappearance of regular regions of phase space already at very small integrability-breaking perturbations. Thus, one may naively expect that quantum many-body systems should display ``typical" relaxation, irrespective of the initial conditions. In this work we demonstrate that this intuition is incomplete. A mixed classical phase space  leaves an imprint on quantum dynamics  in interacting many-body systems, giving rise to slow, atypical thermalization for certain initial conditions. Our starting point is the time-dependent variational principle (TDVP)~\cite{dirac_1930}, which we use to project quantum dynamics onto a classical nonlinear dynamical system. In what follows, we consider both 1D systems, where we choose the variational manifold to consist of translation-invariant matrix-product states (MPS) \cite{Haegeman} with a fixed number of degrees of freedom~\cite{Bernien2017,wenwei18}, as well as higher dimensional systems where we use the infinite tensor tree states (TTS)~\cite{PhysRevB.77.214431}. We demonstrate that the resulting dynamical systems generally have mixed phase space.

Regular regions of phase space are governed by stable periodic trajectories, such as the one in Fig.~\ref{Fig:Z3-PS} below. These periodic orbits have a vanishing Lyapunov exponent, but may be classified according to their ``quantum leakage'', i.e.,  a measure of discrepancy between TDVP and exact quantum dynamics, which  intuitively corresponds to the irreversible entanglement growth. On the one hand, short periodic trajectories, with low quantum leakage, generate  many-body revivals in quantum dynamics. As a proof of principle, we identify a trajectory with a 3-site unit cell that gives rise to robust revivals, and also generalize the notion of scars to a 2D square lattice of Rydberg atoms. On the other hand, even the trajectories with  \emph{high} leakage out of the variational manifold leave a measurable signature on the quantum system: initializing the quantum system in the vicinity of these trajectories significantly slows down thermalization compared to generic initial conditions. We independently confirm these findings by numerical simulations based on exact diagonalization and iTEBD~\cite{Vidal07}.

Similar to the known examples of few-body systems~\cite{Bohigas}, we expect that regular regions of mixed phase space  give rise to non-thermal eigenstates in the many-body quantum system. Thus, mixed phase space would provide a more general mechanism for slow thermalization compared to quantum scars. We establish TDVP as a practical method for finding models with non-ergodic quantum dynamics in \emph{arbitrary} spatial dimensions, which complements its recent applications to quantum thermalization~\cite{Altman17,Green18}.  
Our results also open the door to generalizing the results on few-body chaos to many-body systems, and for approaching the KAM  theorem in quantum systems by utilizing the classical KAM for the TDVP equations of motion. This is distinct from other approaches that rely on broken Bethe-ansatz integrability~\cite{Konik15} or the absence of quantum resonances in the MBL phase~\cite{Serbyn13-1,Huse13,ImbriePRL}.

The rest of the paper is organized as follows. In the remainder of this section, we briefly introduce the TDVP and the tensor network states of interest. In Sec.~\ref{Sec:mixed}, we show how mixed phase space emerges in the TDVP equations that describe the dynamics of Rydberg atom chains, recently realized experimentally~\cite{Schauss2012,Labuhn2016,Bernien2017}. We demonstrate that TDVP allows to identify a stable trajectory that gives rise to \emph{new} quantum revivals beyond those that were probed in recent experiments~\cite{Bernien2017}. In Sec.~\ref{Sec:2D} we show that higher dimensional bipartite lattices of Rydberg atoms also display mixed phase space and the associated quantum revivals. In Sec.~\ref{Sec:leak} we show that quantum leakage can be used to distinguish trajectories that lead to quantum revivals, and to find deformations of the model that \emph{improve} the revivals originating from a given periodic orbit. In Sec.~\ref{Sec:Ising}, we study the transverse-field Ising model  (TFIM) in a longitudinal field, which is a typical example of a thermalizing system. In this case we show that mixed phase space does not give rise to many-body revivals, but leads to a state-dependent thermalization rate. Finally, we conclude with the discussion of open directions and outlook in Sec.~\ref{Sec:Discuss}.
\begin{figure*}[t]
\begin{center}
\includegraphics[width=1.99\columnwidth]{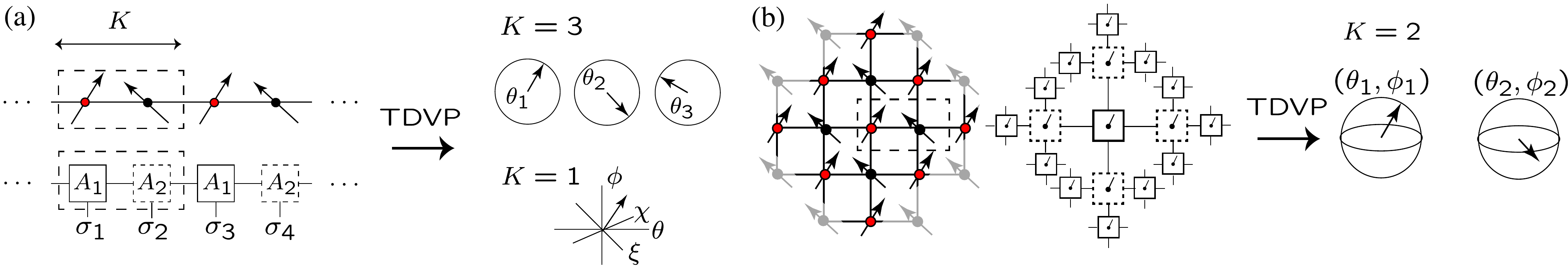}\\
\caption{ \label{Fig:cartoon}  (a) The state of a quantum system that is short-range entangled and translationally invariant with a unit cell of size $K$ can be described by an MPS with the same size of unit cell. (b) The wavefunction of the square lattice is approximated by a tensor tree state (TTS) in the form of a Cayley tree with connectivity $C=4$. The state with $K=2$ unit cell is described by a tree where two different tensors alternate. 
}
\end{center}
\end{figure*}

\subsection{A brief overview of TDVP}

TDVP equations of motion are obtained by extremizing the action~\cite{kramer1981geometry},
\begin{equation}\label{Eq:action}
S = \int dt\, \mathcal{L} ,
\quad 
\mathcal{L}  =  \frac{i}{2}(\braket{\psi|\dot{\psi}}-\braket{\dot{\psi}|\psi})-\braket{\psi|H|\psi}.
\end{equation}
Such action yields a set of nonlinear classical  equations of motion~\cite{kramer1981geometry},
\begin{equation}\label{Eq:TDVPMPS}
\sum_{a}\dot{{x}}_{a}\text{Im}\braket{\partial_{{x}_{b}}\psi|\partial_{{x}_{a}}\psi} = -\frac{1}{2}\partial_{{x}_{b}}\braket{\psi|H|\psi},
\end{equation}
where the variational parameters are real valued, $x_{i} \in \mathbb{R}$. The equations of motion can also be obtained by minimizing the discrepancy between exact quantum dynamics and its projection onto the variational manifold, see Appendix~\ref{App:eom} for a detailed derivation.

TDVP has been succesfully applied to the cases where the variational manifold is spanned by MPS~\cite{ueda2006least,Haegeman} as well as finite tensor tree states (TTS)~\cite{bauernfeind2019time}. Below we introduce the basics of MPS  parametrization, which will be used to describe the dynamics of the Rydberg atom model in Sec.~\ref{Sec:mixed} and the Ising model in~Sec.\ref{Sec:Ising}. In addition, we discuss TTS parametrization that will be used in Sec.~\ref{Sec:2D} to capture the dynamics of the Rydberg model generalized to higher dimensions.

\subsection{Matrix product states and tensor tree states}

The simplest tensor network state is the matrix product state. The manifold of translationally-invariant MPS for a system of size $L$ with a unit cell of fixed size $K$ is defined as follows [see Fig.~\ref{Fig:cartoon}(a)]:
\begin{multline}\label{Eq:psi-A}
\ket{\psi(\{{\bf x}_i\})}
=
\sum_{\{\sigma\}}\Big(V^L
\prod_{m=0}^{L/K-1}\left[A^{\sigma_{1+mK}}({\bf x}_{{1}})
 A^{\sigma_{2+mK}}({\bf x}_{2}) 
 \ldots\right.
 \\
 \left.
 A^{\sigma_{K+mK}} ({\bf x}_{K}) \right]
 V^R\Big)
  \ket{\sigma_1}\ket{\sigma_2}\ldots \ket{\sigma_{L}},
\end{multline}
where $A^\sigma(\bf x)$ is $\chi\times\chi$ matrix ($\chi$ is the bond dimension) that depends on $N$ real on-site variational parameters,  $\vec x =(x_1,\ldots, x_N)$. The physical indices $\sigma$ label the basis of on-site Hilbert space, which we take to be a spin-1/2 degree of freedom with $\sigma=\uparrow,\downarrow$. The index $m=1,\ldots, L/K$ labels the unit cells, and in what follows we take the thermodynamic limit $L\to\infty$. The state in Eq.~(\ref{Eq:psi-A}) is assumed to be normalized, and $V^{L,R}$ denote some boundary vectors the choice of which is not important for our purposes. 

The MPS state~(\ref{Eq:psi-A}) has $N$ parameters for each of the $K$ sites within its unit cell. Hence, one needs to specify overall $NK$ real parameters, $x_a$ with $a=1,\ldots,NK$, to fully fix the state.  Below we will use the MPS ansatz restricted to a unit cell of small size and with a small number of variational parameters. These restrictions allow us to obtain \emph{analytical} equations of motion, making the analysis more transparent. For larger number of parameters, similar analysis could be performed numerically. 

As another extension, a close relative of the MPS ansatz are the TTS with higher connectivity -- see Fig.~\ref{Fig:cartoon}(b). On the one hand, TTS allow to mimic the topology of two-dimensional and higher-dimensional lattices. On the other hand, the absence of loops still allows for an analytical derivation of equations of motion. In this work we restrict ourselves to TTS which approximate bipartite lattices with a natural choice of two-site unit cell. The manifold of TTS with two inequivalent sites, for a system of size $L$  and connectivity $C$, is defined by generalizing Eq.~(\ref{Eq:psi-A}) to the case where $A^\sigma(\bf x)$  is a rank-$C$ tensor, with bond dimension $\chi$. The product of matrices in Eq.~(\ref{Eq:psi-A}) is extended to contraction of indices on all bonds connecting tensors according to Fig.~\ref{Fig:cartoon}(b). While TTS with fixed bond dimension cannot capture the loop effects of higher dimensional lattices, they can still describe the local physics of weakly entangled states. In Sec.~\ref{Sec:2D} we find TTS ansatz to  be very accurate in capturing the quantum dynamics of certain initial states in dimensions $d>1$.

\section{Mixed phase space in TDVP dynamics in one dimension \label{Sec:mixed}}

In this Section, we investigate the phase portrait of the classical nonlinear TDVP equations for an MPS manifold of low bond dimension. Below we introduce the model and variational state, and then demonstrate the mixed nature of phase space and its effects on quantum dynamics.

\subsection{PXP model\label{Sec:PXP-intro}}
\begin{figure*}[t]
\begin{center}
\includegraphics[width=1.99\columnwidth]{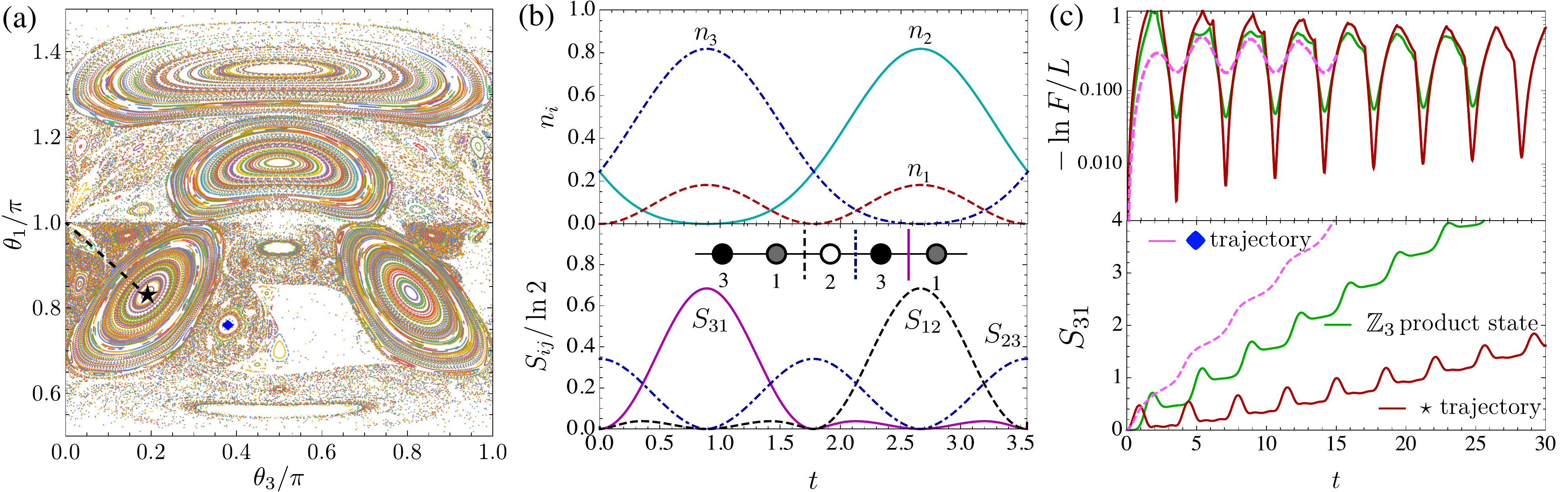}
\caption{ \label{Fig:Z3-PS}
(a) Poincar\'e sections at $\theta_2=0$ for a three-site unit cell reveal mixed phase space with large regular regions. A stable periodic orbit that gives rise to robust revivals is denoted by $\star$ symbol, while $\color{blue}\blacklozenge$ denotes a longer orbit that does not give rise to revivals in exact dynamics. Dashed line represents a particular cut through the regular region, discussed in Sec.~\ref{sec:regular}.
(b) Variational TDVP prediction for the local density of the Rydberg excitation (top panel) and the entanglement entropy (bottom panel) for the shortest-period trajectory labelled by $\star$ on the Poincar\'e section in (a). Inset shows the unit cell choice and entanglement cuts.
{(c) Quantum dynamics for $|\mathbb{Z}_3\rangle$ initial state and two entangled initial states marked in (a) obtained from  iTEBD with $\chi=900$. Top panel: decay of fidelity per spin has regular minima that correspond to revivals of fidelity in finite size system.  Bottom panel: entanglement entropy. The stable ``$\star$'' trajectory features a better fidelity revival, as well as strongly suppressed entanglement growth. Growth of entanglement limits iTEBD simulation to shorter times for $|\mathbb{Z}_3\rangle$, $\color{blue}\blacklozenge$ states.}
}
\end{center}
\end{figure*}
In this section we focus on the PXP model~\cite{Bernien2017}, describing a 1D chain of Rydberg atoms defined by the Hamiltonian:
\begin{equation}\label{Eq:PXP}
 H_\text{PXP}  = \sum_{i=1}^L P_{i-1}\sigma^x_i P_{i+1},
\end{equation}
where $\sigma_x$ is the standard Pauli matrix, and operator  $P_j = (1-\sigma_j^z)/2$ projects to $\downarrow$ state on site $j$. The projectors encode the kinetic constraint due to the Rydberg blockade: two neighboring atoms are not allowed to be simultaneously excited. Unless explicitly stated otherwise, we work in the thermodynamic limit, $L\to\infty$, and restrict the Hilbert space to spin configurations without two adjacent up spins, $|\ldots{\uparrow}{\uparrow}\ldots\rangle$. This is the largest connected component of the full Hilbert space for the Hamiltonian in Eq.~(\ref{Eq:PXP}). The PXP model is interacting and non-integrable~\cite{Turner2017}, yet its relaxation strongly depends on the initial conditions~\cite{Bernien2017}. For instance, there is fast thermalization when the system is prepared in a down-polarized state, $\ket{\downarrow\downarrow\ldots}$, while other initial states, such as  $|\mathbb{Z}_2\rangle = \ket{\uparrow\downarrow\uparrow\downarrow\ldots}$  state, exhibit revivals of local observables~\cite{Bernien2017}, entanglement entropy~\cite{Turner2017}, and even the many-body wave function~\cite{Turner2018}. 

Below, we apply  the ansatz in Eq.~(\ref{Eq:psi-A}) with $K=3$ to PXP model. We parametrize the matrices $A^\sigma$ by two angles, ${\bf x}_i = (\theta_i, \phi_i)$:
\begin{equation}\label{Eq:A-matrix}
 {A}^\uparrow(\theta_{i}, \phi_{i}) = 
 \left(\begin{matrix}
 0 &  i e^{-i\phi_{i}}\\
 0 & 0
 \end{matrix} \right),
 \
  { A}^\downarrow(\theta_{i}, \phi_{i}) = 
 \left(\begin{matrix}
 \cos\theta_{i}  &  0\\
 \sin\theta_{i}& 0
 \end{matrix} \right).
\end{equation}
The matrix $A^\uparrow \propto \sigma^+$ satisfies the condition $A^\uparrow (\theta, \phi)A^\uparrow (\theta', \phi')=0$, thus effectively imposing the constraint that no two adjacent spins are in $\ket{\uparrow}$ state. 

The Hamiltonian in Eq.~(\ref{Eq:PXP2D}) has particle-hole symmetry in the spectrum: $E\to -E$, since it anti-commutes with ${\cal C}  = \prod_i \sigma^z_i$. In this case the system's equations of motion have a class of solutions with phase angles being stationary, $\phi_{i}=0$, which corresponds to dynamics restricted to a \emph{flow-invariant subspace} in the language of dynamical systems. This subspace arises from the presence of particle-hole symmetry and time-reversal invariance of the PXP Hamiltonian in Eq.~(\ref{Eq:PXP}), see Appendix~\ref{App:eom}. Note that $\corr{H} = 0$ vanishes identically when  $\phi_{i}=0$, thus imposing no additional constraint. Restricting to $\phi_{i}=0$ subspace for $K=2$ unit cell, Ref.~\cite{wenwei18} found an unstable periodic trajectory in a two-dimensional flow invariant subspace. This trajectory is  intimately connected with the revivals from  the initial Ne\'el state $\ket{\mathbb{Z}_2}=\ket{\downarrow\uparrow\downarrow \uparrow\ldots}$.

 However, a two-dimensional phase space is non-generic and does not allow for chaos: in two dimensions any periodic trajectory fragments the phase space into dynamically isolated regions. Thus, we focus on ans\"atze with 3 or more parameters, which have regular and chaotic regions coexisting in phase space. We will show below that such behavior, known as  ``mixed phase space'', is robust and persists for various deformations of the PXP model. We will then investigate the implications for quantum dynamics, finding new initial states that exhibit revivals and slow thermalization. 

\subsection{Revivals in a three-site unit cell \label{Sec:PXP-Z3} }

Higher-dimensional variational phase space naturally arises when unit cell size is increased, while  particle-hole symmetry is kept intact. This describes the dynamics of initial states that are periodic in space, with period $K=3$ (or more) lattice sites. An experimentally relevant example of such initial states is $|\mathbb{Z}_3\rangle = |{\uparrow\downarrow\downarrow}\ldots\rangle$, which can be experimentally prepared in a Rydberg system with larger blockade radius~\cite{Bernien2017}. The resulting equations of motion for $\theta_{i}$, $i=1,2,3$, obtained from the TDVP projection of the PXP model can be found in Appendix~\ref{App:Z3}. 

In order to analyze the dynamical flow, we consider its Poincar\'e section. The flow generates a discrete mapping known as the Poincar\'e map~\cite{strogatz} that maps a given point $(\theta_1,\theta^*_2,\theta_3)$ on the chosen hyperplane $\theta_2 = \theta_2^*$ to a position  $(\theta'_1,\theta^*_2,\theta'_3)$ where the trajectory intersects this plane again. Periodic trajectories correspond to stationary points of the Poincar\'e map, while in the case of chaotic behavior the system returns to the same plane at a location that is generally far away from the previous encounter. Successively iterating the Poincar\'e map for many different initial conditions yields the Poincar\'e section, shown in Fig.~\ref{Fig:Z3-PS}(a). 

Fig.~\ref{Fig:Z3-PS}(a) reveals the phase portrait characteristic of dynamical systems with mixed phase space.  Cross-sections of at least four large stable KAM tori are evident, surrounded by a number of smaller tori and chaotic regions. The analysis of periodic orbits reveals a stable shortest-period orbit denoted by $\star$  in Fig.~\ref{Fig:Z3-PS}(a) and  located at $(\theta_1,\theta_2,\theta_3)=(0.8356\pi, 0, 0.1923\pi)$. Other large tori surround the orbits related to this particular orbit via symmetry transformations. 

Next, we analyze the local observables and entanglement entropy of the $\star$ periodic orbit from Fig.~\ref{Fig:Z3-PS}(a). Fig.~\ref{Fig:Z3-PS}(b) shows the time evolution of the entanglement entropy for the variational solution, which  indicates that the dynamics on this orbit never passes through a product state. To see this, note that when the entanglement for one cut, say $S_{12}$, vanishes, the entanglement for a different cut is non-zero, e.g., $S_{23}>0$. Interestingly, this short-range entanglement improves the fidelity of revivals in the corresponding quantum system. Indeed, the closest product state to this trajectory is $|\mathbb{Z}_3\rangle = |{\downarrow\uparrow\downarrow}\ldots\rangle$. However, preparing the quantum system in a slightly entangled state positioned on the periodic TDVP orbit gives rise to more robust revivals.  The improvement of revivals is illustrated in Fig.~\ref{Fig:Z3-PS}(c). {In the top panel, we show the negative logarithm of fidelity,
\begin{eqnarray}\label{eq:fid}
 F(t) = |\langle \psi_0 | \exp(-i H t)| \psi_0\rangle |^2,
\end{eqnarray}
normalized by the system size, $-(\ln F)/L$, that quantifies the fidelity decay per spin.
The minima  in this quantity correspond to the best fidelity revivals of all finite-size subsystems. These minima display a regular pattern, with smaller values (corresponding to higher return amplitude) when the system is initialized on the TDVP trajectory compared to $|\mathbb{Z}_3\rangle$ product state.} {The bottom panel of Fig.~\ref{Fig:Z3-PS}(c) shows that the growth of entanglement is notably slower for the initial state positioned on the TDVP orbit, consistent with better fidelity revivals}.

\subsection{Effect of regular regions of phase space}\label{sec:regular}

Above we considered the dynamics of a quantum system initialized on the periodic orbit. However, any such classically stable orbit is surrounded by a regular region of phase space where TDVP dynamics is constrained to a torus. In order to investigate the effect of this region on quantum dynamics, we consider a family of initial states $\ket{\psi(c)}$, where the parameter $c\in[0,1]$ interpolates along the dashed line in Fig.~\ref{Fig:Z3-PS}(a) between the periodic orbit and the point $\theta_{1}=0, \theta_3=\pi$.  

\begin{figure}[t]
\begin{center}
\includegraphics[width=0.99\columnwidth]{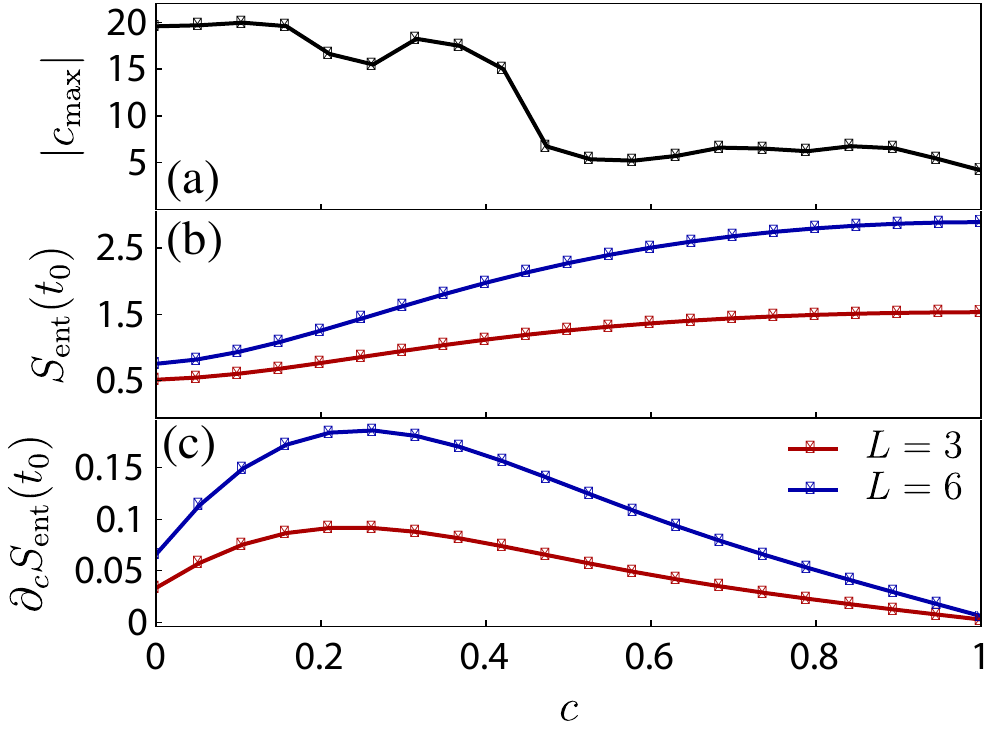}\\
\caption{ \label{Fig:mixed} 
(a) Rapid drop in the largest component of the Fourier spectrum of a local observable, $n_1(t)$, signals the onset of chaotic behavior.  (b) 
Entropy of a region of size $L$ at fixed time $t_0=15$ remains low when the quantum dynamics is launched near the center of the regular region (we use iTEBD algorithm to simulate quantum dynamics). When the border of the regular region is approached, the value of entanglement rapidly increases with changing the initial state, defined by the value of $c$. When $c$ approaches one, the dynamics becomes insensitive to the changes in the initial state due to rapid thermalization. (c) The derivative of entropy along the trajectory is most sensitive around the boundary of the regular region. 
}
\end{center}
\end{figure}

Launching classical dynamics for different initial conditions, we clearly see the termination of the regular region in the Fourier spectrum of local observables, Fig.~\ref{Fig:mixed}(a). Inside the regular region, the Fourier spectrum is strongly peaked. In contrast, upon entering the chaotic sea when $c\approx 0.42$, one observes a rapid drop in the amplitude of the maximal Fourier harmonic component that signals a transition to chaotic dynamics with the continuous Fourier spectrum. For the same set of initial conditions, quantum dynamics shows  a crossover between non-ergodic behaviour, with slow entanglement growth, and thermalizing dynamics, as witnessed by the behavior of entanglement entropy, see Fig.~\ref{Fig:mixed}(b).  The changes in the behavior of the quantum system are much more smooth: intuitively, one may think of a quantum system initialized in a finite region of the phase space, with the size of this region being determined by quantum fluctuations.  This smoothens out any abrupt transition between the non-ergodic and thermalizing behaviors. Nevertheless,   Fig.~\ref{Fig:mixed}(c) shows that the maximal sensitivity of quantum dynamics to a change in initial conditions is achieved near the border of the regular region.

In addition to entanglement entropy, we observed similar behavior of inverse participation ratios (IPR) of local density matrices, measured at the time when the dynamics has saturated (not shown). The IPRs are found to be low in the regular region, consistent with the expectation that the system does not explore the full configuration space when launched there. In contrast, the participation ratios rapidly increase near the border of the regular region and attain thermal values outside of it. 

The classical dynamical system also has many longer orbits surrounded by ``thinner'' tori -- for example, see the orbit marked by the blue diamond in Fig.~\ref{Fig:Z3-PS}(a). We observe that these orbits do not give rise to long-lived oscillations in exact dynamics of the PXP model, and they are characterized by rapid entanglement growth -- see the dashed line in the bottom panel of Fig.~\ref{Fig:Z3-PS}(c). Since numerous \emph{stable} orbits appear in the Poincar\'e sections, we need to understand which of these orbits give rise to strong quantum revivals in exact quantum dynamics. We will return to the question of how to distinguish orbits in Sec.~\ref{Sec:leak}, after first considering the dynamics and mixed phase space on higher-dimensional lattices in the next section.

\section{Revivals and mixed phase space in higher dimensions \label{Sec:2D}}

In this section we introduce the PXP model in higher dimensions and apply the TTS ansatz  to capture its dynamics. We will focus on a 2D square lattice, and compare the predictions of the TTS dynamics against exact diagonalization.
We will also discuss generalizations of these results to any bipartite lattice in an arbitrary number of spatial dimensions. 

\subsection{PXP model in higher dimensions\label{Sec:2D-state}}

We start by introducing a constrained model on the square lattice, inspired by the 1D PXP model:
\begin{equation}\label{Eq:PXP2D}
 H_\text{PXP}  = \sum_{i,j}  P_{i-1,j}P_{i,j-1}\sigma^x_{i,j}P_{i+1,j}P_{i,j+1}+\mu_z  n_{i,j},
\end{equation}
where the pair of indices $i,j$ is used to label lattice sites. In addition to the PXP term, we have also include a possible perturbation in terms of chemical potential $\mu_z$, which couples to the Rydberg excitation density operator $n=(1+\sigma^z)/2$. The presence of four projectors allows a given atom to get excited only if all neighboring atoms are in the ground state.  Similarly to the PXP model in 1D, we will restrict our attention to the largest connected subspace of the Hilbert space, where no adjacent atoms are excited.  

After generalizing the quantum model to the 2D case, we turn to the variational state that can capture the quantum dynamics. Inspired by the ans\"atze considered above for the 1D lattice, we introduce tree tensors $A^\sigma_{i_{1},\ldots, i_4}$, which forbid excitations on adjacent sites. Relegating the detailed discussion to Appendix~\ref{App:tree}, here we list only the non-zero elements of the tensor $A^\sigma$ that is parametrized by two angles, $(\theta,\phi)$:
\begin{equation}\label{Eq:TTS}
\begin{split}
A^{\downarrow}_{i_{1},i_2, i_{3},0} &= \cos^{I}{\theta}\sin^{3-I}{\theta},\\
A^{\uparrow}_{0,\ldots 0,1} &= i e^{-i \phi},\\
\end{split}
\end{equation}
where $i_{1,\ldots 4}$ are virtual indices, and  $I = \sum^{3}_{j=1} i_{j}$ counts the number of ``excitations'' on the three legs of the tensor. The fourth leg of the tensor  tensor produces an ``output $1$-bit'' on the last virtual index, $i_4$, if the corresponding site is excited and $0$ if it's in the ground state. In the Appendix~\ref{App:tree} we give the form of the ansatz for general connectivity $C$ of the tree. When $C=2$, which corresponds to the 1D limit, this ansatz reproduces Eq.~(\ref{Eq:A-matrix}).

Projection of quantum dynamics generated by Eq.~(\ref{Eq:PXP2D}) onto the TTS manifold with a 2-site unit cell results in the dynamical system of four variables that come in two canonically conjugate pairs, $(\theta_1,\phi_1)$ and $(\theta_2,\phi_2)$. The general form of equations of motion reads:
\begin{subequations}\label{Eq:Z2-dyn}
\begin{eqnarray}\label{Eq:Z2-dyn-a}
 \dot\theta_1 & =&  -\sin{\theta_{1}}\cos^C{\theta_{1}} \cos{\phi_{1}}  \tan{\theta_{2}}- \cos^{C-1}{\theta_{2}} \cos{\phi_{2}},\\ 
  \nonumber \dot\phi_1 & =&  \mu_{z} 
  -
  C\tan\theta_{1}\cos^{C-1}\theta_{2}  \sin\phi_{2}\\\nonumber \label{Eq:Z2-dyn-b}
    & +& \frac{1}{2} \cos^{C-1}\theta_{1}\cot{2 \theta_{2}} \sin{\phi_{1}} (3 + C - (C-1) \cos{2 \theta_{1}}) \\
    &&      -(C-1) {\cos^{C-1}\theta_{1}\sin^2\theta_{1}\sin\phi_{1}}{ \sin^{-1} 2 \theta_{2}},
\end{eqnarray}
\end{subequations}
where the square lattice case (ignoring loops) corresponds to $C=4$. The equations for $\dot\theta_2,\dot\phi_2$ can be obtained by substitution $1\leftrightarrow2$. In addition, the energy density,
\begin{equation}\label{Eq:Ham-Z2}
\frac{\corr{H}}{L}  =\frac{\mu_{z} \sin^2\theta_{2}+\cos^{C-1}{\theta_{1}}\sin{2\theta_{2}}\sin{\phi_{1}}}{2+2\cos^2\theta_{2}\tan^2\theta_{1}}+1\leftrightarrow 2
\end{equation}
is a conserved quantity, provided the system satisfies the equations of motion, Eqs.~(\ref{Eq:Z2-dyn}). The existence of a conserved quantity effectively reduces the dimensionality of the phase space to three dimensions, as dynamics is restricted to constant energy surfaces.  

\begin{figure}[tb]
\begin{center}
\includegraphics[width=0.985\columnwidth]{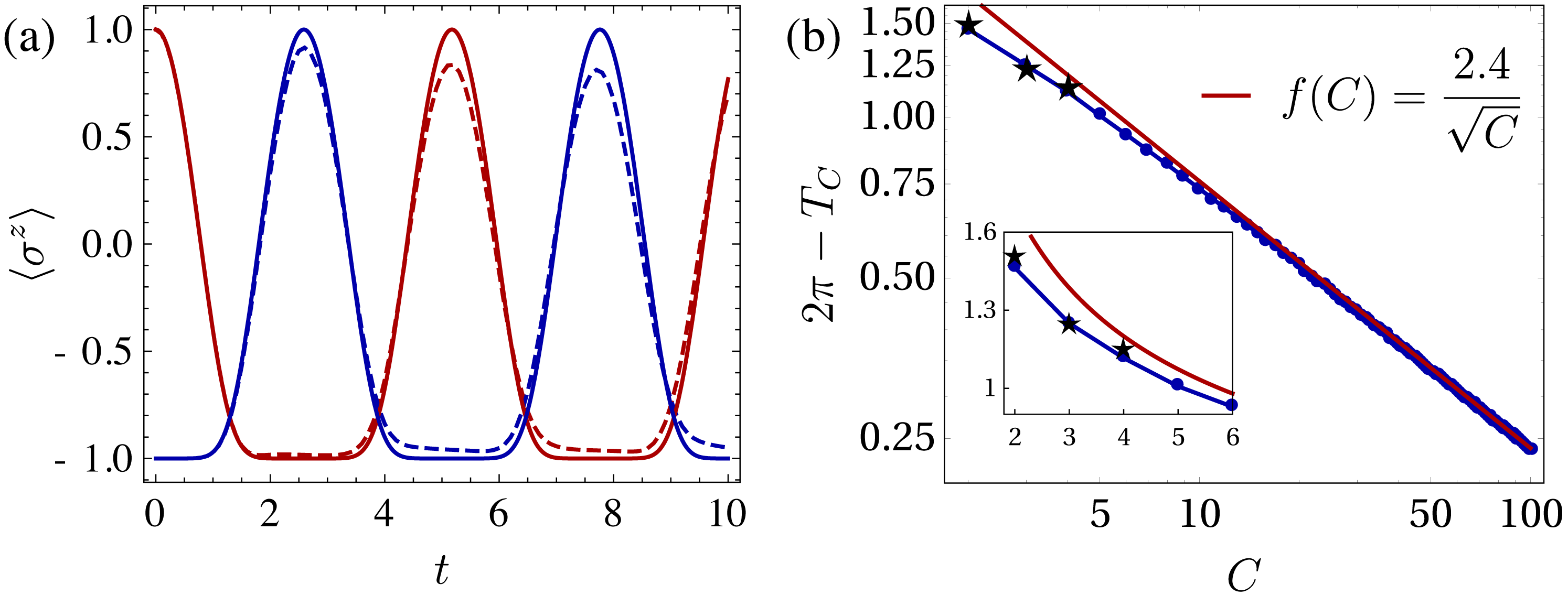}
\caption{ \label{Fig:Z2-2D}(a) TDVP dynamics on a tree (solid lines) compares favorably with exact dynamics on a $ 4\times 4$ lattice (dashed lines). Different colors denote different sites in the unit cell. (b) Scaling of the period  of the isolated periodic orbit as a function of the connectivity of the TTS. Stars indicate exact results for chain, honeycomb and square lattices.
}
\end{center}
\end{figure}

\begin{figure*}
\begin{center}
\includegraphics[width=1.99\columnwidth]{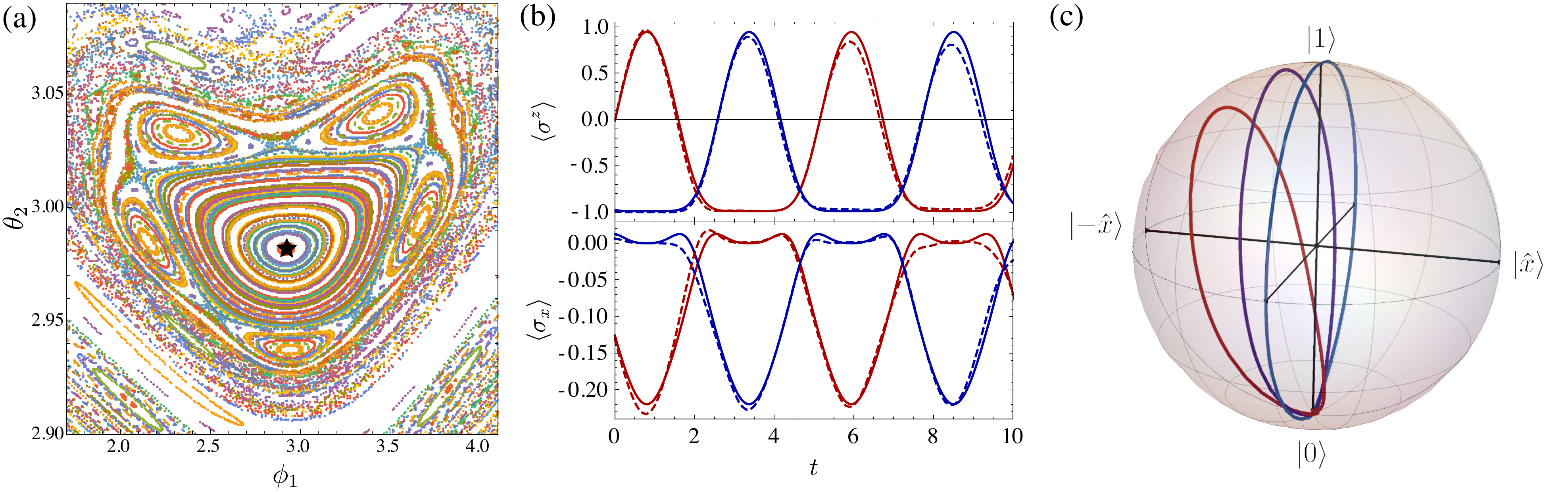}\\
\caption{ \label{Fig:Z2-PS}
(a) Poincar\'e section for $\mu_z = 0.225$ at $(\theta^*_1= 1.25 \pi,\dot \theta_1<0)$ and $\corr{H}=0$ reveals mixed phase space with large regular regions. 
The Poincar\'e section is obtained by initializing the system on a grid of points with the step $\delta \theta_2=\delta \phi_2 =8 \cdot 10^{-3}$ and numerically integrating the evolution up to a time $t=3000$ using \texttt{NDSolve} routine in Mathematica.
(b) Comparison between TDVP predictions for the dynamics of local observables ($\mu_z=0.225$) on the periodic trajectory (solid lines) and the exact dynamics of a $4 \times 4$ lattice with periodic boundary conditions. 
(c)  Bloch sphere visualization of the individual spin dynamics on the TDVP trajectory. The trajectory on the big meridian corresponds to the case $\mu_z=0$. Non-zero chemical potential $\mu_z=0.225,\ 0.65$ causes a progressive tilting of the trajectory and the development of a knot around the south pole. 
}
\end{center}
\end{figure*}

\subsection{Generalizing scars to arbitrary bipartite lattices}

When $\mu_z = 0$ and one restricts to a flow-invariant subspace $\phi_{1,2}=0$, Eqs.~(\ref{Eq:Z2-dyn}) reduce to a two-dimensional flow. This flow has an isolated periodic trajectory that generalizes the trajectory found in Ref.~\cite{wenwei18} to an arbitrary bipartite lattice. This trajectory is responsible for the revivals of a  state in which all atoms on the first sublattice are in the Rydberg state, $\uparrow_1$ and all atoms on the second sublattice are in the ground state, $\downarrow_2$. In Fig.~\ref{Fig:Z2-2D}(a) we show that quantum dynamics of a local observable $\braket{\sigma^z}$ on the $4\times 4$ square lattice agrees well with TDVP dynamics based on $C = 4$ TTS. Moreover, we compare the TDVP dynamics of a wavefunction satisfying the kinetic constraints and having the same weights as the tree parametrization, see Eq.~(\ref{Eq:weights}) in App.~\ref{App:TTS}, to the tree dynamics and find perfect agreement (not shown). The fact that TDVP dynamics that takes into account loops agrees with TTS TDVP demonstrates that presence of loops is not important for local observables.

As connectivity increases, the isolated periodic orbit of the TTS manifold is expected to become more accurate, and effectively exact at $C \rightarrow \infty$. In this limit, the sites that are in their ground state cannot flip unless all their neighbours are also in their ground state. The periodic trajectory can be understood by studying the $C\rightarrow \infty$ limit of Eq.~(\ref{Eq:Z2-dyn}). At the first quarter of the period, the excited sites on the first sublattice relax $\uparrow_1\to\downarrow_1$ while their neighbors remain in the ground state due to infinite connectivity. At the second quarter of the period, the second sublattice undergoes the transition $\downarrow_2\to \uparrow_2$. The half period over which two sublattices exchange their state is given by $\pi$ in our notations, so the total period of the orbit is $T_{C\to \infty} = 2 \pi$. Figure~\ref{Fig:Z2-2D}(b) shows that period of the orbit approaches its $C\to \infty $ limit approximately as $T_C = 2 \pi - 2.4/\sqrt{C}$.

\subsection{Phase space and revivals in higher dimensions}

When detuning $\mu_z\neq 0$, the dynamics generated by Eq.~(\ref{Eq:Z2-dyn}) takes place on constant energy surfaces in four-dimensional space. In order to visualize such dynamics, we fix $\theta_1$ variable and show the resulting Poincar\'e sections in Fig.~\ref{Fig:Z2-PS}(a) [the precise value of $\theta_1$ is not important, and we have chosen $(\theta^*_1 = 1.25 \pi, \dot \theta_1<0)$].   Energy conservation results in complicated surfaces in the space $(\phi_1,\theta_2,\phi_2)$ that cannot be globally projected. Therefore, we show a small part of the Poincar\'e section that can be projected onto $(\phi_1,\theta_2)$ plane. We find that the point $(\phi_1,\theta_2)\approx(2.921,2.981)$ is a stationary point of the Poincar\'e map that corresponds to a stable periodic trajectory.  We observe that this periodic trajectory is surrounded by circle-shaped contours which are the intersections of KAM tori by our section plane. 

A key question concerns the correspondence between regular variational dynamics and exact quantum dynamics. Fig.~\ref{Fig:Z2-PS}(b) shows this comparison for local observables $\langle \sigma_i^{x,z} \rangle$, for the system initialized on the periodic trajectory (indicated by $\star$ in Fig.~\ref{Fig:Z2-PS}). It demonstrates that the classical periodic trajectory gives rise to pronounced quantum revivals; moreover, the exact quantum dynamics agrees with the TDVP predictions up to $t\lesssim 10$. We note that the slight disagreement at $t= 0$ is caused by the difference between the structure of a TTS (that cuts loops and introduces redundant degrees of freedom) and true many-body wave function on a two-dimensional lattice (see App.~\ref{App:TTS1D}).
In order to visualize the variational (classical) dynamics, Fig.~\ref{Fig:Z2-PS}(c) shows the evolution of the trajectory of an individual spin $(\corr{\sigma^x_1(t)},\corr{\sigma^y_1(t)},\corr{\sigma^z_1(t)})$ on the Bloch sphere as perturbation $\mu_z\neq 0$ is turned on. The second spin in the unit cell performs similar oscillations shifted in phase by $\pi$.

 For $\mu_z=0$ the trajectory gives  oscillations of the spin between $\uparrow$ and $\downarrow$ states, and the expectation value $\corr{\sigma^x(t)}$ is zero at all times. In agreement with physical intuition, non-zero chemical potential acts as a ``magnetic field'' in  the $z$-direction, tilting the plane in which the spin rotates. Surprisingly, in addition to the tilt, the trajectory slightly winds around the south pole of the Bloch sphere.  This winding reflects the fact that we are dealing with an interacting system rather than with a precession of independent spins, which shows that this orbit cannot be obtained as a smooth deformation of the unstable periodic orbit found in the case of $\mu=0$. 

In the entire range of $\mu_z\in (0,0.65]$ the regular regions remain pronounced in the phase space of the system in Eqs.~(\ref{Eq:Z2-dyn}), thus demonstrating the persistence of mixed phase space. The deformation of the trajectory away from the poles upon increasing $\mu_z$ suggests that the most stable revivals occur for initial states that are different from the Ne\'el product state, $\mathbb{Z}_{2}$, by having some amount of entanglement.

\section{Characterizing trajectories by leakage \label{Sec:leak}}

Variational dynamics of the PXP model and its deformations, studied above, revealed  mixed phase space with multiple stable orbits. The Lyapunov exponent, which is often used to quantify chaos in TDVP~\cite{Green18} and semiclassical~\cite{Scaf} dynamics, vanishes for all these orbits and thus cannot be used to distinguish them. In this section we show that \emph{quantum leakage} can be used instead as a heuristic for distinguishing  trajectories that give rise to revivals in exact quantum dynamics.  

\subsection{Rate of leaving the variational manifold\label{Sec:leak-def}}

TDVP approach accurately approximates short-time quantum evolution of initial states in the chosen MPS manifold, but at long times the errors will grow because exact quantum dynamics generally brings the system out of the variational manifold.  Intuitively,  the TDVP error for a given initial state is linked to the rate of entanglement growth for that state, since MPS with low bond dimension can only represent weakly entangled states. When the entanglement grows significantly, the state requires an MPS description with a larger bond dimension that lies outside the variational manifold. Making this connection mathematically precise is challenging and is beyond the scope of this work. Instead, here we  characterize the error by quantum leakage defined as the instantaneous rate at which the exact quantum wave function leaves the variational manifold~\cite{Haegeman,wenwei18}:
\begin{equation}\label{Eq:leak2}
\gamma^2(\{x_a\}) = \lim_{L\to\infty}\frac{1}{L}|| (iH + \dot{x}_b \partial_{x_b}) \ket{\psi(\{x_a\})} ||^2.
\end{equation}
In Appendix~\ref{App:leak} we discuss the computation of this quantity in the TDVP framework. This calculation involves the square of the Hamiltonian operator, thus effectively it contains information that goes beyond the TDVP equations of motion. The normalization factor is chosen such that $\gamma$ assumes a finite value in the thermodynamic limit. 

In the previous section we demonstrated that TDVP equations of motion may have periodic trajectories, ${x}_a(t) = {x}^{(0)}_a(t)$, with a period $T$.  If the quantum system were following the TDVP dynamics exactly, that would imply persistent oscillations in local observables and perfect revivals of the many-body quantum fidelity, i.e., $F(t=nT)=1$ in Eq.~(\ref{eq:fid}), where the initial state can be any MPS wave function that belongs to the periodic trajectory, $|\Psi\rangle = |\psi({x}^{(0)}_a(t))\rangle$. The disagreement between quantum dynamics and TDVP will generally preclude such perfect revivals (except for models with perfect scars~\cite{Soonwon2018}). However, we conjecture that one can obtain a \emph{lower bound} on the many-body fidelity revival using quantum leakage.

Assuming an extensive scaling of the fidelity, $F(T) = e^{- f_TL}$ at some fixed short time and $L\to \infty$ limit, and using the large system size limit of Eq.~(\ref{Eq:leak2}), we posit the following upper bound on $f_T$,
\begin{equation}\label{Eq:fid-bound}
 f_T  
 \leq 
 \Gamma_T, 
 \quad
 \text{with}
\quad
\Gamma_T= \left[\int_0^T dt\, \gamma(\{x^{(0)}_a(t)\})\right]^2,
\end{equation}
that translates to a lower bound on the fidelity. This bound can be justified in the limit of small leakage for a finite size system, see Appendix~\ref{App:leak}. However, its extension to the thermodynamic limit is non-trivial and at present it remains a conjecture.  Below we present  a specific example which demonstrates the usefulness of quantum leakage and provides a test of the bound.

\subsection{Leakage and revivals criterion\label{Sec:improve}}

\begin{figure}[t]
\begin{center}
\includegraphics[width=0.999\columnwidth]{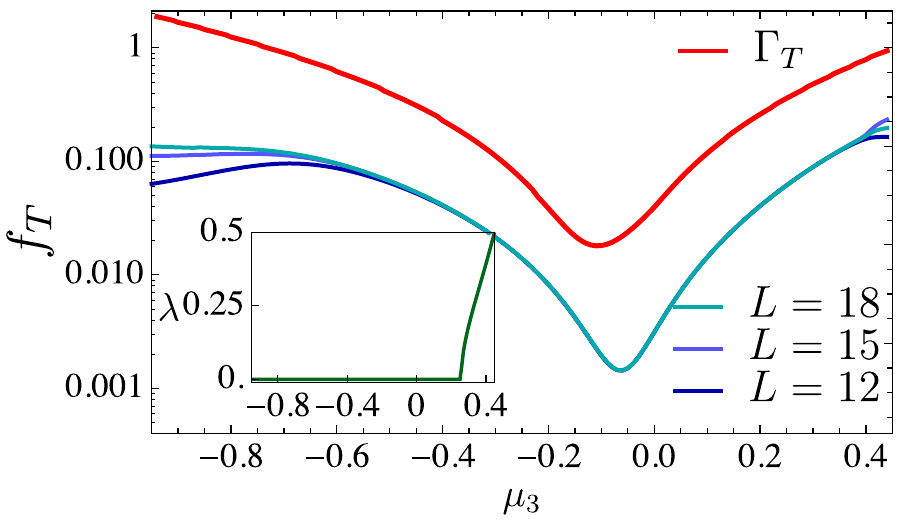}\\
\caption{ \label{Fig:Z3-chaos}
  (a) TDVP quantum leakage presents an upper bound on  $f_T$. Despite not being tight, this bound captures the qualitative effect of the applied perturbation. The data illustrates that $f_T$ is system size independent, and therefore is well defined for $\mu_3\in[-0.6,0.4]$. The inset shows that the Floquet exponent $\lambda$ only becomes non-zero for $\mu_3> 0.255$, suggesting that there is no direct relation between $\lambda$ and leakage. Finally, for  $\mu_3\geq0.45$ the periodic trajectory disappears.}
\end{center}
\end{figure}

We study the evolution of the Poincar\'e sections and corresponding trajectory under the deformation of the PXP model that makes it more thermalizing. Specifically, we consider the following deformation,
 \begin{equation}\label{Eq:h3}
 H_{\mu_3} = 
\mu_3 \sum_i P_{i-2}\left(\sigma^+_{i-1}\sigma^-_i\sigma^+_{i+1} +\sigma^-_{i-1}\sigma^+_i\sigma^-_{i+1}\right) P_{i+2},
\end{equation}
which is the lowest order term that induces spin flips (expected to facilitate thermalization) while maintaining particle-hole symmetry. 

The presence of particle-hole symmetry in $H=H_\text{PXP}+H_{\mu_3}$ allows us to use the same ansatz as in Sec.~\ref{Sec:PXP-Z3} with three variational parameters $\theta_{1,2,3}$. The resulting TDVP equations of motion are again cumbersome and can be found in Appendix~\ref{App:Z3}. Analysis of these Poincar\'e sections reveals a pronounced asymmetry between the effect of this deformation for $\mu_3=\pm 0.25$. The deformation with $\mu_3<0$ increases the size of regular regions in the phase space, see Fig.~\ref{Fig:deformed-PS}. In contrast,  positive $\mu_3$ causes the regular regions of the phase space to shrink in size and the phase portrait looks progressively more chaotic (Fig.~\ref{Fig:tra-analyze}). Exact diagonalization shows a similar asymmetry between positive and negative~$\mu_3$: for $\mu_3>0$ thermalization, diagnosed via level repulsion and average eigenstate entanglement entropy, becomes more pronounced.

In order to quantify the effect of the deformation, we use the expression from Appendix~\ref{App:leak} for $\gamma(\{x_a\})$ to calculate the integrated leakage,  $\Gamma_T$, defined in Eqs.~(\ref{Eq:leak2})-(\ref{Eq:fid-bound}), along the trajectory.  Fig.~\ref{Fig:Z3-chaos} shows that the leakage attains a minimum at a small negative value $\mu^\text{TDVP}_3\approx -0.11$. For larger negative values of $\mu_3$ the leakage starts to increase again. Although the leakage overestimates the suppression of the fidelity revivals, the exact diagonalization results for $f_T = -\ln F(T)/L$ follow qualitatively the same trend as  $\Gamma_T$. The $f_T$ obtained from exact quantum dynamics achieves a minimum at $\mu_3^\text{ED} \approx -0.06$, which corresponds to the best fidelity revivals. 
The optimal value of the deformation,  $\mu_3^\text{ED}$, is approximately twice smaller compared to the TDVP prediction,  $\mu^\text{TDVP}_3$. Nevertheless,  TDVP may be used to find the best operators suitable for stabilizing quantum many-body revivals~\cite{Soonwon2018}. Moreover, in contrast to the revivals from $|\mathbb{Z}_2\rangle$ state~\cite{wenwei18}, our results show that generally one must account for the effect of the deformation on the initial state.

Eventually, for large absolute values of $\mu_3$, we observe that the quantity $f_T$ is no longer well defined, since the data for different system sizes do not collapse onto each other in Fig.~\ref{Fig:Z3-chaos}.  We attribute this to fast thermalization at such deformation parameters. At these values, quantum dynamics no longer exhibits many-body revivals, and simultaneously $\Gamma_T$ is on the order of unity. Thus, we propose
\begin{eqnarray}
\Gamma_T \sim 1
\end{eqnarray}
as a tentative criterion for when a periodic trajectory ceases to result in revivals. This criterion was previously heuristically conjectured in Ref.~\cite{wenwei18} in the context of dynamics of $Z_2$ state. With this criterion, it is natural to test other periodic trajectories visible in the Poincar\'e sections of the TDVP dynamics. In Appendix~\ref{App:PXP-trajectories} we present such an analysis for $\mu_3=0.25$, which shows that the majority of trajectories have leakage that is larger than one, thus not giving rise to quantum revivals. 

\subsection{Floquet exponents and quantum leakage} \label{Sec:lyap}

We demonstrated the utility of quantum leakage in classifying trajectories and predicting which trajectories give rise to fidelity revivals. The leakage gives a particular measure of instability of the dynamics with respect to \emph{external} degrees of freedom -- it quantifies how quickly quantum evolution leaves the variational manifold. At the same time, the TDVP equations of motion can be classified by an \emph{internal} measure of instability of the TDVP dynamics. To that end,  in the case of chaotic motion, one uses the Lyapunov exponent, whereas in the case of periodic orbits the Floquet exponents provide a conceptually similar but reparametrization-invariant measure~\cite{ChaosBook}, see App.~\ref{App:PXP-trajectories} for details. 

It is natural to ask if the Floquet exponent and quantum leakage are directly related to each other. The answer is negative: Fig.~\ref{Fig:Z3-chaos}(inset)  shows how a non-zero Floquet exponent $\lambda$ emerges for $\mu_3\geq 0.255$. However, the non-zero value of $\lambda$ does not have any influence on the leakage and this unstable trajectory still gives rise to fidelity revivals for $\mu_3 \in (0.255,0.4]$. This example also suggests that mixed phase space provides a more generic mechanism for weak ergodicity breaking. Stable trajectory in the center of a regular region of mixed phase space that gives rise to ``regular eigenstates''~\cite{Percival,Bohigas} may become unstable upon a deformation of the Hamiltonian, leading to the appearance of quantum scars~\cite{Turner2017,wenwei18} which are quantum eigenstates affected by the unstable trajectories.

\section{ Non-universal thermalization \label{Sec:Ising}}

Above we focused on periodic trajectories that arise in classical TDVP equations, with small quantum leakage. These trajectories give rise to spectacular revivals of the fidelity and slow thermalization (or even its absence, as is the case for the deformed PXP model~\cite{Soonwon2018}). In this Section we consider the opposite situation where quantum leakage is strong.  In this regime, TDVP fails to capture the dynamics of local observables beyond times of order $O(1)$. Nevertheless, we show that  mixed phase space still leaves an imprint on quantum dynamics, leading to \emph{non-universal} thermalization.

\subsection{Transverse-field Ising model \label{Sec:Ising-intro}}

We study the quantum Ising model with transverse and longitudinal fields (TFIM), a popular model for investigating eigenstate thermalization,  defined by a Hamiltonian,
\begin{equation}\label{Eq:TFIM}
H_\text{TFIM} = \sum_i J_z \sigma^z_i\sigma^z_{i+1}+ h_z \sigma^z_i + h_x\sigma^x_i,
\end{equation}
with fixed values of parameters $(J_z, h_z,h_x) = (1,0.4,1)$. Thermalization in this model was established for somewhat different parameters $(J_z, h_z,h_x) = (1,0.8090,0.9045)$ in Ref.~\cite{Huse14}; we choose a smaller value of $h_z$ to facilitate the TDVP analysis. {Very similar behavior is also observed for other values of couplings, $h_x=0.9$ and $1.1$.}  
  
In order to capture the dynamics, we use the following MPS ansatz with bond dimension $\chi=2$,  
\begin{subequations}
\label{Eq:TFIM-ansatz}
\begin{eqnarray}
A^\uparrow &=& 
\left(\begin{matrix}
  \cos\theta\cos\xi e^{i\chi/2}&  \cos\theta\sin\xi e^{-i \chi/2}\\
0&0
 \end{matrix} \right)
 \\
 A^\downarrow &=& 
\left(\begin{matrix}
0&0\\
  \sin\theta \sin\xi e^{i(\phi-\chi/2)}& \sin\theta \cos\xi e^{i(\chi/2+\phi)}
 \end{matrix} \right)
\end{eqnarray}
\end{subequations}
that depends on four parameters, $(\theta, \phi,\xi, \chi)$, per spin. Such a choice of the ansatz is inspired by ``trotterization'' of the evolution operator  $e^{-i H_\text{TFIM}\delta t} \approx e^{-i H_{0}\delta t}e^{-i H_{1}\delta t}$, where $H_{0}$ and $H_{1}$ are the transverse-field term and remaining terms in $H_\text{TFIM}$, respectively. We restrict ourselves to translation-invariant states, setting the unit cell size $K=1$. We note the such a choice of tensors does not result in a normalized wave function. Hence, the derivation of the equations of motion is more complicated and we follow the general procedure outlined in Appendix~\ref{App:TFIM}. We perform the numerical integration of the equations of motion and obtain the phase portraits of the TDVP dynamics. 

\subsection{Dependence of dynamics on the initial state \label{Sec:Ising-ent}}

We consider the TDVP and exact quantum dynamics for initial states specified by the ansatz in Eq.~(\ref{Eq:TFIM-ansatz}), with fixed energy density $\corr{H}/L = 0.19$. This corresponds to initial states at high but finite temperature in the thermodynamic limit. In this case, the TDVP dynamics reveals the presence of mixed phase space (see Fig.~\ref{Fig:TFIM} in Appendix~\ref{App:TFIM2}). However, the integrated leakage for the shortest stable trajectory with the period $T= 2.097$ is close to one, $\Gamma_T = 0.57$. Fully consistent with such values, the trajectory does not result in oscillations of the fidelity and all local observables relax by the time of order $t\sim 2$. 

\begin{figure}[t]
\begin{center}
\includegraphics[width=0.95\columnwidth]{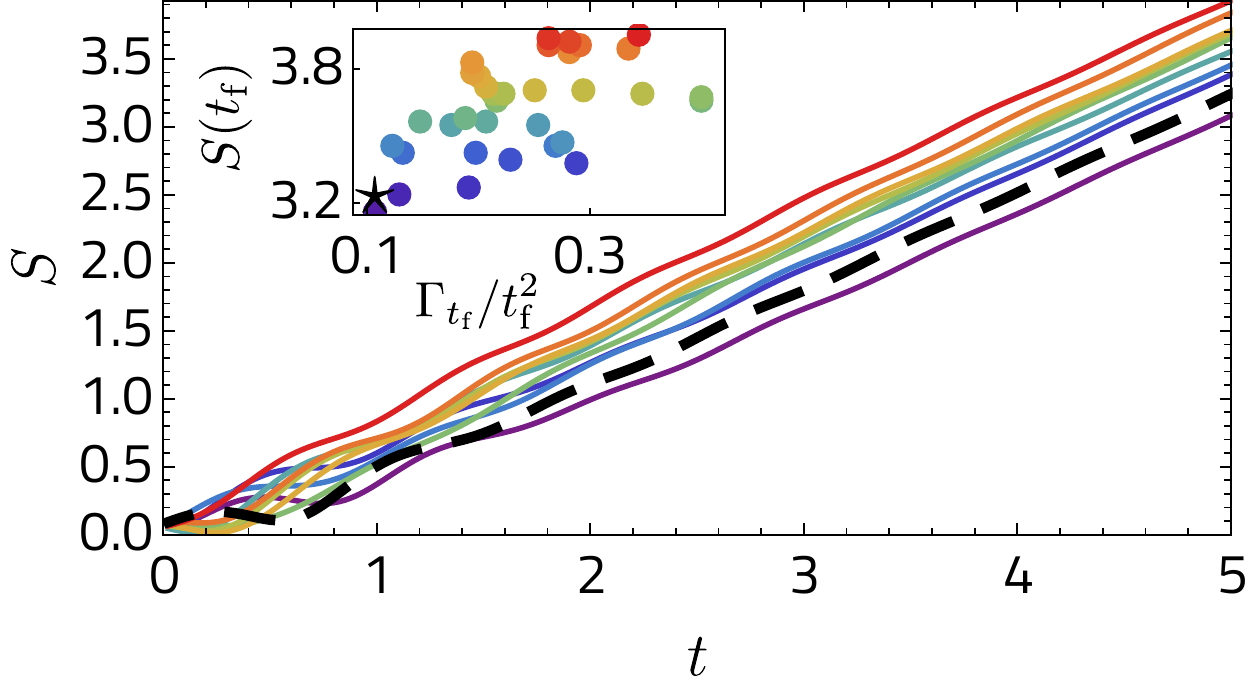}\\
\caption{ \label{Fig:thermo-Ising}
Bipartite entanglement dynamics for different MPS initial states shows strong dependence of entanglement growth on the initial state. The inset reveals that the amount of entanglement at  time $t_\text{f}=5$ is correlated with the MPS leakage averaged over TDVP dynamics, $\Gamma_{t_\text{f}}/t_\text{f}^2$, where the periodic trajectory is shown by $\star$. Using initial conditions on the periodic trajectory results in slow entanglement dynamics (thick dashed line), but there are initial conditions that give even slower entanglement growth. The data is obtained with iTEBD with a maximum truncation error $\sim 10^{-10}$.}
\end{center}
\end{figure}

In order to study the effect of the periodic trajectory, we consider an ensemble of initial states specified by the ansatz in Eq.~(\ref{Eq:TFIM-ansatz}), with the same energy density and a small initial value of the entanglement entropy $S_\text{ent} \in[(2/3) S_{0},(4/3)S_{0}]$, where $S_{0} \sim 0.08$ is the minimal value of entanglement entropy of the given periodic orbit. 
Since all these states have the same energy density, nearly the same initial value of the entanglement, and are translationally invariant, we expect that after  short time sufficient to equilibrate local observables, the entanglement dynamics should be independent of the initial state.
Fig.~\ref{Fig:thermo-Ising} shows that these expectations are not correct, and the entanglement at late times is still strongly influenced by the vicinity to the periodic orbit.  At the same time, all these initial conditions result in the same saturation value of entanglement for finite subsystems, see Fig.~\ref{Fig:TFIM-sat}. 

 Thus we conclude that these initial conditions not only have different short-time behavior of the entanglement, but are also characterized by a \emph{different velocity} of the entanglement spreading. In Appendix~\ref{App:TFIM2} we demonstrate that, for the present choice of parameters, the eigenstates of this model have properties fully consistent with ETH, hence such dynamics cannot be explained by the existence of anomalous eigenstates. We note that a similar phenomenon was observed in a Floquet version of TFIM with somewhat weaker integrability-breaking field~\cite{Prosen19}. 

The correlation between TDVP leakage integrated over a short time and the entanglement spreading rate -- see Fig.~\ref{Fig:thermo-Ising}(inset) -- provides further support for the relevance of  TDVP dynamics  for entanglement spreading even at late times. The statistical analysis gives a correlation coefficient of 0.53 with $p$-value of 0.0007, suggesting that correlations are statistically significant. Appendix~\ref{App:PXP-deform} reveals similar phenomenology in the case of a strongly deformed PXP model, where the different velocity of entanglement spreading and correlation between leakage and entanglement is even more apparent. The existence of such a correlation suggests that it may be possible to use the leakage as an upper bound on the value of entanglement at long times. 

Finally, we note that different rates of entanglement spreading observed in a thermalizing system is well-known in the context of integrable models~\cite{Calabrese05,Alba,Calabrese18}. In the latter case the dependence of entanglement dynamics on the initial state stems from the presence of multiple conserved quantities. In the present case we are dealing with a non-integrable TFIM; however, it is tempting to conjecture the existence of one (or a few) approximately conserved quantities related to regular regions of the phase space. Also, a possible relation between such entanglement spreading and the existence of slowly growing operators found in the TFIM~\cite{Huse15E} remains an intriguing open question. 
 
\section{Discussion\label{Sec:Discuss}}

To summarize, we examined the relation between exact quantum dynamics and an effective classical non-linear system, obtained by projecting quantum dynamics via TDVP onto the restricted MPS manifold. This approach is different from the time-dependent mean field  (which can be viewed as a particular case of $\chi=1$ MPS) and semiclassical treatments used in  few-body quantum chaos in that it incorporates short-range entanglement. We demonstrated the relevance of mixed phase space, identified  in the TDVP dynamics, for the exact dynamics of quantum many-body systems. We used quantum leakage to distinguish two qualitatively different situations:
\begin{enumerate}
\item[(i)]  In the \emph{small-leakage} regime, stable periodic trajectories lead to long-time revivals of the many-body fidelity and oscillations of local observables. This provides a more general mechanism of weak ergodicity breaking compared to quantum many-body scars in an arbitrary number of spatial dimensions.

\item[(ii)] In the \emph{strong-leakage} regime, exactquantum dynamics does not follow the TDVP predictions, but the mixed nature of phase space influences the rate of entanglement growth. 
\end{enumerate}

In the weak leakage case, (i), we establish the TDVP method as an indispensable tool in searching for new stable periodic trajectories leading to fidelity revivals. This is demonstrated in Secs.~\ref{Sec:mixed} and \ref{Sec:2D} by finding revivals in one-dimensional PXP model and its higher-dimensional generalizations.  In addition to revivals, the entire regular region of the phase phase space was shown to have influence on quantum dynamics. Moreover, this regular region  is expected to give rise to special \emph{regular eigenstates} in the many-body spectrum (we use the term ``regular eigenstates'' to distinguish the atypical eigenstates that originate from the regular regions in the phase space from ``scars'' that come from unstable periodic trajectories). The detailed investigation of these states is left to future work. Next, in Sec.~\ref{Sec:leak} we demonstrated that the revivals in the PXP model are much more robust to deformations of the Hamiltonian than was previously reported in the literature~\cite{Turner2017,Turner2018}. However, in the course of the deformation one has to follow the periodic trajectory that is influenced by the deformation, which can be inferred from TDVP. 

In the case of strong leakage, (ii), considered in Sec.~\ref{Sec:Ising}, the thermalizing TFIM is also shown to have mixed phase space. Initializing the system in the regular region of the phase space near  the stable periodic trajectory yields a slower rate of entanglement growth compared to other states with the same energy density. These results show that the low bond dimension TDVP, despite its failure to capture rapid entanglement growth, can be useful in searching for slowly thermalizing initial conditions in interacting quantum systems. 

It is important to keep in mind that TDVP, in general, gives a multitude of possible ways to map a quantum many-body system onto a classical dynamical system, see Fig.~\ref{Fig:cartoon}.  While some TDVP ans\"atze may find unstable trajectories, thus motivating the term ``quantum scars''~\cite{wenwei18}, more generic embeddings may result in mixed phase space. Hence it remains an open question which of the atypical eigenstates observed in different models~\cite{Turner2017,wenwei18,Turner2018,Khemani2018,Iadecola19,lin2018exact,Bernevig2017,BernevigEnt,Titus19,Pretko19,Nandkishore19,Pollmann19,Papic19} correspond to many-body scars, and which to regular eigenstates.

In addition to establishing the methodological utility of low bond dimension TDVP as a diagnostic of anomalous thermalization, our results pose questions related to the consequences of mixed phase space for the spectrum, level statistics and other indicators of thermalization in quantum many-body systems.  As we mentioned in the introduction, mixed phase space is known to influence the semiclassical limit of  few-body quantum systems~\cite{Bohigas}. In this case an intermediate spectral statistics is known to arise~\cite{Berry84,Prosen_1993}, and the system has an increased stability with respect to noise when it is prepared in the regular region of phase space~\cite{HaakeBook}. Our work suggests that TDVP can be used to generalize these results to the case of quantum many-body chaos. However, we note that this generalization may be highly non-trivial as quantum many-body eigenstates in the zero momentum sector may see imprints from mixed phase space in TDVP ans\"atze  with different unit cell choices, as in Fig.~\ref{Fig:cartoon}. These studies are of practical importance, as they can result in ways of delaying thermalization and increasing the stability of a quantum many-body system to noise. Thus it is interesting to explore different mechanisms that can lead to low-leakage trajectories. Two prospective directions include weakly perturbed mean-field models (e.g. TFIM with small $J\ll h_x,h_z$) and models with constrained Hilbert spaces. In addition, it is interesting to check for the effect of mixed phase space in other methods used to treat interacting systems, such as Gutzwiller projected dynamics~\cite{Marco} and MPS-based DMFT~\cite{DFTMPS}. 

In a complementary direction,  TDVP was recently proposed as a method to capture thermalization. In particular, Ref.~\cite{Altman17} demonstrated that physical properties saturate in the case of TFIM when  bond dimension $\chi>2$. In addition, Ref.~\cite{Green18} studied the spectrum of Lyapunov exponents in the case of high bond dimension TDVP. Our work shows that in the case of low bond dimension one may encounter non-chaotic behavior in the TDVP dynamics, limiting the utility of the Lyapunov exponent in such cases. On the one hand, one may expect that  mixed phase space does not persist for large bond dimensions, as increasing  $\chi$ increases the dimension of the phase space where the TDVP dynamics occurs, making it more susceptible to chaos. On the other hand, for local Hamiltonians the Lieb-Robinson bound~\cite{LiebRobinson} suggests that a small bond dimension is sufficient to capture quantum dynamics at early times. Thus, it is important to understand how the mixed phase space evolves upon including additional MPS parameters that describe longer-range entanglement. Specifically, one can embed one of \mbox{small-$\chi$} MPS ans\"atze considered here into an MPS with a larger bond dimension and track the behavior of the variational parameters that correspond to longer-range entanglement. 

At the more phenomenological level, recent works established numerous examples of anomalous thermalization. In particular, these include slow thermalization emerging from confinement~\cite{Calabrese16,Konik_Robinson2018}, the existence of ``slow operators'' in thermalizing models~\cite{Huse15E}, dynamical phase transitions~\cite{Heyl13}, dependence of dynamics on the initial state in Floquet systems~\cite{Prosen19}, mixed phase space in quantum maps~\cite{Sondhi19}, etc. It would be interesting to understand if all or some of these phenomena can be interpreted using the framework developed here and its straightforward extension to Floquet systems.

Finally, TDVP formulated in the MPS basis could provide a novel pathway to a quantum-mechanical analogue of the KAM theorem. The robustness of quasi-local integrals of motion in MBL phases~\cite{Serbyn13-1,Huse13,ImbriePRL} 
provides an established particular case of the quantum KAM~\cite{AbaninRev}. In a different setting, Ref.~\cite{Konik15} demonstrated the existence of quasi-conserved quantities in  Bethe-ansatz integrable systems deformed by an integrability-breaking perturbation.
These works argue for a quantum KAM without making reference to its classical counterpart. By contrast, our results suggest that one may use classical KAM as a route towards its quantum counterpart. Indeed, recent work~\cite{Soonwon2018} conjectured that a  weak deformation of  the PXP model may lead to perfect many-body revivals in the thermodynamic limit. It remains an open question if such a deformation of the PXP Hamiltonian can allow quantum dynamics to \emph{exactly} follow the TDVP trajectory. More broadly, searching for deformations of quantum models that reduce quantum leakage in the TDVP dynamics or yield integrable TDVP dynamics may provide a specific route to quantum KAM that is expressed in terms of the KAM for the corresponding classical system.

\acknowledgments
We acknowledge useful discussions with E. Altman, B. Budanur, P. Calabrese, A. Green, F. Pollmann, and T. Prosen. The work of D.A. is supported by the Swiss National Science Foundation. C.J.T. and Z.P. acknowledge support by EPSRC grants EP/P009409/1, EP/R020612/1 and EP/M50807X/1. Statement of compliance with EPSRC policy framework on research data: This publication is theoretical work that does not require supporting research data.

\appendix

\section{Variational ans\"atze, equations of motion, and leakage\label{App:eom}}

In this Appendix we provide details on the derivation of TDVP equations of motion used in the main text. We begin by setting up the general framework that gives an efficient way of obtaining the equations of motion. Next, we apply this framework to derive equations of motion for the (deformed) PXP model and TFIM which are analyzed in the main text.  

\subsection{General framework}

\subsubsection{Notations}

In what follows we use TDVP formulated for a wave function that is represented in MPS form. We note that this formulation also includes dynamical mean-field theory as a particular case if one restricts to the product-state form for the variational wave function. While the general theory of TDVP in MPS manifolds was obtained in Ref.~\cite{Haegeman}, here we aim to provide a simpler recipe which can be used to obtain \emph{analytical form} of equations of motion for states with a small number of variational parameters.

We consider translationally-invariant matrix-product state (MPS) ansatz for the wave function, with the size of the unit cell being fixed at $k$, see Eq.~(\ref{Eq:psi-A}) and Fig.~\ref{Fig:cartoon}. Tensor $A^\sigma_{\alpha\beta} ({\bf x})$ has one physical index, $\sigma=\uparrow,\downarrow$ for spin-1/2 degrees of freedom, and two virtual indices $\alpha,\beta = 1,\ldots \chi$, where $\chi$ is the (fixed) bond dimension. This tensor depends on a set of $N$ real variational parameters, $\vec x \in \mathbb{R}^N$. While our Hamiltonian is translationally invariant, we consider initial states that break translational symmetry. More specifically, we allow for initial states with a unit cell of size $K$. Thus, the complete set of variational parameters consists of a  set $\{\vec x_i\}$ with $i=1,\ldots K$, where $i$ labels sites within the unit cell. In order to simplify the notations, in the remainder of Appendix we suppress the dependence of $A$ on $\vec x_i$,  $A_i \equiv A(\vec x_i)$. 

In order for the state $\ket{\psi}$, Eq.~(\ref{Eq:psi-A}), to have system-size independent norm, its unit cell transfer matrix must be non-degenerate with the largest eigenvalue $\lambda_\text{max} = 1$. The unit cell transfer matrix is obtained from a product of individual one-site transfer matrices,
\begin{equation}\label{Eq:tm1}
(T_i)_{\alpha\gamma, \beta\delta} =
\sum_\sigma\ \raisebox{-0.29in}{\includegraphics[width=0.15\columnwidth]{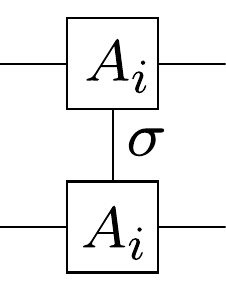}}\
=
 \sum_\sigma (A^\dagger_i)^\sigma_{\alpha\beta} (A_i^\sigma)_{\gamma\delta},
\end{equation}
over the entire unit cell:
\begin{equation}\label{Eq:transfer}
 T_\text{u.c.} 
 =
 T_1\cdot T_2\cdot \ldots \cdot T_K.
\end{equation}
In the above notations all transfer matrices operate in the $\chi^2$-dimensional ``double'' virtual space labeled by a pair of indices [e.g., $\alpha\gamma$ in Eq.~(\ref{Eq:tm1})]. In what follows, vectors in the physical Hilbert space are denoted as $\ket{*}$ and vectors of the double virtual space by $|*)$. The dominant left/right eigenvectors of the unit-cell transfer matrix $T_\text{u.c.}$ that correspond to an eigenvalue with maximal norm, $\lambda_\text{max}=1$ are labelled as $|L)$ and $|R)$.  The boundary tensors $v,u$ in Eq.~(\ref{Eq:psi-A}) do not affect the calculations in the thermodynamic limit as long as $(v^{\dag}v|R) \neq 0$, $(L|u u^{\dag}) \neq 0$.

\subsubsection{Action and gauge choice}
The TDVP equations are obtained by extremizing the action~(\ref{Eq:action})~\cite{Haegeman}. The symmetric choice of the Lagrangian is invariant under time-dependent unitary transformation $\ket{\tilde \psi}=U\ket{\psi}$, where $U = e^{i a(t)}$ is a pure phase. Such a phase is fixed by requiring the system to satisfy the ``mean" Schr\"odinger equation,  
\begin{equation}\label{Eq:phase}
\begin{split}
\braket{\tilde\psi|\dot{\tilde\psi}} &= - i\braket{\tilde\psi|H|\tilde\psi}\\
\Rightarrow \frac{d a }{dt} &=-\braket{\psi|H|\psi} + i\braket{\psi|\dot{\psi}},
\end{split}
\end{equation}
where the phase $a(t)$ is real since the norm of the wave function is time-independent. In what follows we show that this choice of global phase effectively removes all disconnected correlations --- see Eq.~(\ref{Eq:connectedleak}) --- from both the equations of motion and the error, Eqs.~(\ref{Eq:TDVPMPS}) and (\ref{Eq:leak}). 

In the case when the MPS is sparsely parametrized and satisfies the normalization condition, it is often possible to calculate analytically $\braket{\psi|\dot{\psi}}$ and $\braket{\psi|H|\psi}$. In such cases the calculation of equations of motions is greatly simplified compared to Ref.~\cite{Haegeman}. Below we concentrate on such a case, and derive the general form of equations of motion.

\subsubsection{Hamilton's Equations of Motion}\label{TDVP_formalism}

In the case when the MPS manifold is parametrized by complex parameters, the complex conjugate pairs, $z_i$ and $\bar z_i$ can be interpreted as momenta and coordinates. Since we use an explicitly real parametrization, we also separate the $N K$ parameters $\{\vec x_i\}$ into $N K/2$ coordinate variables $\{\bm\theta_i\}$ and $N K/2$ phase variables $\{\bm\phi_i\}$, where $N$ is the number of real on-site parameters that is assumed to be even. The amplitude  variables satisfy the condition of vanishing overlap between the wave function and tangential vector in any of $\bm \theta$ direction:
\begin{equation}\label{Eq:phase1}
\braket{\psi |\partial_{\theta_{\kappa}}{\psi}} = 0, \quad \forall \kappa \in 1, \ldots, \frac{N K}{2}.
\end{equation}
In contrast, the phase variables have a non-vanishing overlap, 
\begin{equation}\label{Eq:phase2}
-i\braket{\psi |\partial_{\phi_{\kappa}}{\psi}} = f_{\kappa}(\{\bm \theta_i\}),
\end{equation}
characterized by the set of real functions $f_{\kappa}(\{\bm \theta_i\}) \in \mathbb{R}$ that depend solely on amplitudes. It is convenient to define the matrix,
\begin{equation}
\eta_{\kappa \kappa'}=\partial_{\theta_{\kappa'}} f_{\kappa} = 2 \text{Im} \braket{\partial_{\theta_{\kappa'}}\psi|\partial_{\phi_{\kappa}}\psi},
\end{equation}
which can be understood as the imaginary part of the manifold metric, $g_{ i j} = \braket{\partial_{x_{i}}\psi|\partial_{x_{j}}\psi}$.

Extremizing the action, Eq. (\ref{Eq:action}), over phase variables $\{\bm \phi_i(t)\}$ gives equations of motion for $\{\bm \theta\}$,
\begin{equation}\label{eq:thetadot}
\sum_{\kappa'} \eta_{\kappa \kappa'} \dot{\theta}_{\kappa'}=  \partial_{\phi_{\kappa}}\braket{\psi|H|\psi},
\end{equation}
while extremizing the action over amplitude variables yields equations of motion for $\{\bm \phi\}$,
\begin{equation}\label{eq:phidot}
\sum_{\kappa'} \eta^{T}_{\kappa \kappa'} \dot{\phi}_{\kappa'}=  -\partial_{\theta_{\kappa}}\braket{\psi|H|\psi}.
\end{equation} 
The equations (\ref{eq:thetadot})-(\ref{eq:phidot}) are Hamilton's equations in a manifold with the Poisson bracket,
\begin{equation}
\{g,\rho\} = {\partial_{\bm \theta}}{g}\;  \hat{\eta}^{-1}{\partial_{\bm\phi}}\rho-{\partial_{\bm\phi}}\rho\;\hat{\eta}^{-1}{\partial_{\bm\theta}}g,
\end{equation}
thus making apparent their symplectic structure.

The equations of motion obtained above are determined by the expectation value of the Hamiltonian and the set of functions $f_{\kappa}(\{\bm \theta_i\})$. Below we discuss the calculation of these ingredients in the specific case of PXP and TFIM.

\subsection{TDVP for PXP model\label{App:Z2}}

 The MPS ansatz used for PXP model, Eq.~(\ref{Eq:A-matrix}), has a left-canonical form since the matrix $A$ satisfies the condition $\sum_{\sigma}A_{i}^{\sigma\dag}A_{i}^{\sigma}=\mathds{1}$ for any values of $(\theta_i,\phi_i)$. In such a case, all single-site transfer matrices $T_i$ have largest eigenvalue $\lambda_\text{max}=1$, with dominant left eigenvector $(L^{i}|=(1,0,0,1)$. On the other hand, the right dominant eigenvector depends on the size of the unit cell. It should be noted that the left-canonical form of the MPS ansatz is not crucial for the calculation, but it greatly simplifies the analytical expressions. 

\subsubsection{EOMS for PXP with $K=2$ ansatz}

For the two-site MPS ansatz we get the following transfer matrix,
\begin{equation}
T = \left(\begin{matrix}  \cos^2\theta_{1}\cos^2\theta_{2} +\sin^2\theta_{2} & 0 & 0 & \cos^2\theta_{1} \\
 \cos\theta_{1}\cos^2\theta_{2}\sin\theta_{1} & 0 & 0 & \cos\theta_{1}\sin\theta_{1} \\
\cos\theta_{1}\cos^2\theta_{2}\sin\theta_{1} & 0 & 0 & \cos\theta_{1}\sin\theta_{1} \\
  \cos^2\theta_{2}\sin^2\theta_{1} & 0 & 0 & \sin^2\theta_{1} \\ \end{matrix} \right),
\end{equation}
and the corresponding right dominant eigenvector, 
\begin{equation}
|R)
= \cos^2\theta_{2}\tan\theta_{1}({\cos^{-2}\theta_{2}\tan^{-1}\theta_{1}},1,1,\tan\theta_{1}).
\end{equation}
Using the expression for the left dominant eigenvector, we calculate their overlap,
\begin{equation}\label{Eq:LR-Z2}
(L|R) = 1+(\cos{\theta_{2}}\tan{\theta_{1}})^2
\end{equation}
that enters as a normalization factor in all expressions. 

The functions $f_{1,2}(\{\bm\theta\})$ are calculated as follows. We explicitly use translational invariance of the MPS state to obtain
\begin{multline}\label{Eq:corfunc_1d}
f_{1} =-i\corr{\psi|\partial_{\phi_1}\psi}\\
 =- i\frac{L}{2}\ \raisebox{-0.2in}{\includegraphics[width=0.7\columnwidth]{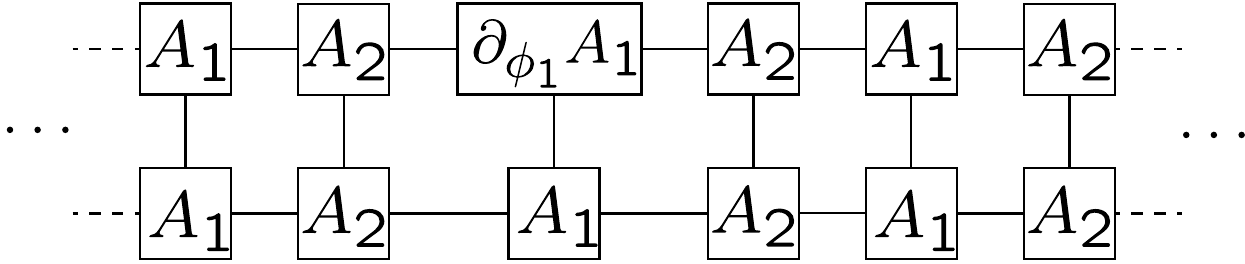}}\ \ 
,
\end{multline}
where the tensor $\partial_{\phi_1} A_1 \equiv \partial_{\phi_1} A^\sigma(\theta_1,\phi_1)$ can be obtained from Eq.~(\ref{Eq:A-matrix}). Replacing the environments by the dominant left and right vectors, we get the following expression 
\begin{equation}
f_1 = -i\frac{L}{2}\frac{(L|T^{\partial_{\phi_1}}_1 T_{2}|R)}{(L|R)},
\end{equation}
where we defined the matrix $T^{\partial_{\phi_1}}$ as 
\begin{multline}
(T^{\partial_{\phi_1}}_1)_{\alpha\gamma, \beta\delta} =
\sum_\sigma\ \raisebox{-0.29in}{\includegraphics[width=0.15\columnwidth]{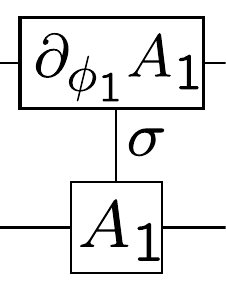}}\
=
 \sum_\sigma (A^\dagger_1)^\sigma_{\alpha\beta} (\partial_{\phi_1}A_1^\sigma)_{\gamma\delta} 
\\
 = \left(\begin{matrix}  0 & 0 & 0 &- i \\
 0 & 0 & 0 & 0 \\
0 & 0 & 0 & 0 \\
  0 & 0 & 0 & 0 \\ \end{matrix} \right).
\end{multline}
Using the explicit form of the matrix $T^{\partial_{\phi_1}}$ we obtain the following expression for $f_1$:
\begin{equation}
f_{1} =  -\frac{L}{2}\frac{\sin^2{\theta_{2}}}{1+(\cos{\theta_{2}}\tan{\theta_{1}})^2}.
\end{equation}
Repeating the calculation for $f_2$ gives us:
\begin{equation}
f_{2} =  -\frac{L}{2}\frac{\cos^2{\theta_{2}}}{\cos^2{\theta_{2}}+\cot^2{\theta_{1}}}.
\end{equation}

The expectation value of the PXP Hamiltonian has two different contributions, depending on which sites the Hamiltonian acts. First we calculate the contraction of the local Hamiltonian term with the immediate environment. This results in matrices operating in the double virtual space,
\begin{multline}\label{Eq:H121}
\mathcal{H}^{1,2,1}_{PXP}
=\ \raisebox{-0.29in}{\includegraphics[width=0.35\columnwidth]{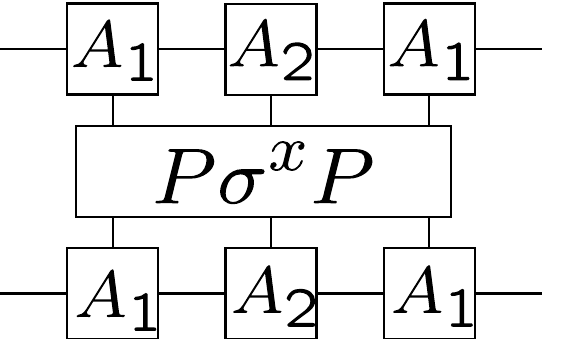}}\
\\=\cos{\theta_{2}}\sin{2\theta_{1}}\sin{\phi_{2}}
\left(\begin{matrix}  \cos ^2\theta_{1} & 0 & 0 & 0 \\
  \sin \theta_{1} \cos \theta_{1} & 0 & 0 & 0 \\
  \sin \theta_{1} \cos \theta_{1} & 0 & 0 & 0 \\
  \sin ^2\theta_{1} & 0 & 0 & 0 \\ \end{matrix} \right).
\end{multline}
The matrix  $\mathcal{H}^{2,1,2}_{PXP}$ where the operator $\sigma^x$ operates on the $A_2$ site can be obtained from Eq.~(\ref{Eq:H121}) by replacing $\theta_{1} \leftrightarrow \theta_{2}$ and $\phi_{1} \leftrightarrow \phi_{2}$. The case of the chemical potential can be treated in a similar way, but the resulting matrices $\mathcal{H}^{1,2}_{\mu_z}$ are independent of variational parameters, 
\begin{multline}\label{Eq:Hmuz1}
\mathcal{H}^{1}_{\mu_z}
=\ \raisebox{-0.29in}{\includegraphics[width=0.13\columnwidth]{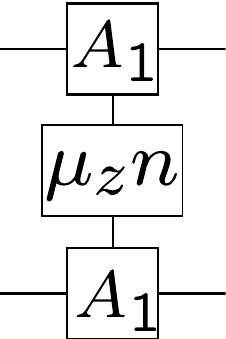}}\
=
\mu_z \left(\begin{matrix}  0 & 0 & 0 & 1 \\
 0 & 0 & 0 & 0 \\
0 & 0 & 0 & 0 \\
  0 & 0 & 0 & 0 \\ \end{matrix} \right)
=\mathcal{H}^{2}_{\mu_z}.
\end{multline}
Now by contracting these expressions with the left and right dominant vectors and taking into account the normalization, we obtain the expectation value of the respective terms in the Hamiltonian, 
\begin{multline}\label{Eq:corrHPXP}
\braket{\psi| H_{PXP}|\psi}
= \frac{L}{2}\frac{(L_\text{max}|\mathcal{H}^{1,2,1}_{PXP}T_{2}+T_{1}\mathcal{H}^{2,1,2}_{PXP}|R_\text{max})}{(L_\text{max}|R_\text{max})}
\\
= \frac{L}{2}\Big(\frac{2\cos^3\theta_{2}\sin\phi_{2}\tan\theta_{1}}{1+\cos^2\theta_{2}\tan^2\theta_{1}}+ 1\leftrightarrow 2 \Big)
\end{multline}
\begin{multline}\label{Eq:corrHmuz}
\braket{\psi| H_{\mu_z}|\psi} =\frac{L}{2} \frac{(L_\text{max}|\mathcal{H}^{1}_{\mu}T_{2}+T_{1}\mathcal{H}^{2}_{\mu}|R_\text{max})}{(L_\text{max}|R_\text{max})}
\\
= \frac{L}{2}\mu_{z}\Big(\frac{\sin^2\theta_{2}}{1+\cos^2\theta_{2}\tan^2\theta_{1}} + 1\leftrightarrow 2 \Big).
\end{multline}
Summing these expressions we obtain Eq.~(\ref{Eq:Ham-Z2}) in the main text. The resulting equations of motion are obtained by substituting the specific form of the functions $f_{1,2}$ and the Hamiltonian density into Eqs.~(\ref{eq:thetadot})-(\ref{eq:phidot}). We note that the system size $L$ enters both the functions $f_i$ and the expectation value of the Hamiltonian. Hence, it gets canceled in the equations of motion. The equations are the same as Eq.~(\ref{Eq:Z2-dyn}) for $C=2$.

\subsubsection{Flow invariant subspace}

As we noted in the main text, Eqs.~(\ref{Eq:Z2-dyn}) possess an additional symmetry when $\mu_z=0$. Namely, $\corr{\psi|H_{PXP}|\psi}$ in Eq.~(\ref{Eq:corrHPXP}) vanishes for any values of $\theta_{1,2}$ when $\phi_{1,2}=0$. This leads to the emergence of a \emph{flow-invariant subspace}: if we start the dynamics from any point with $\phi_{1,2}=0$, the phase variables will remain zero throughout the flow. In other words, the hyperplane $\phi_{1,2}=0$ is preserved under the flow generated by  equations of motion~(\ref{Eq:Z2-dyn}) with $\mu_z=0$ since the time derivative $\dot\phi_{1,2} = 0$ when $\phi_{1,2}=0$, for any values of $\theta_{1,2}$. 

The emergence of the flow invariant subspace $\phi_{1,2}=0$ is not restricted to $K=2$ unit cell. In facts, it is a generic feature of the MPS ansatz~(\ref{Eq:A-matrix}) with any unit cell, provided the Hamiltonian is invariant under particle-hole and time-reversal symmetries (by time-reversal symmetry we mean the invariance of $H$ under  complex conjugation, or equivalently the absence of any terms with an odd number of $\sigma^y$ matrices in the Hamiltonian). Indeed, particle-hole symmetry requires that the operator ${\cal C} = \prod \sigma^z_i$ anticommutes with $H$, or in other words ${\cal C} H{\cal C} = -H$. In addition, if we use the fact that the Hamiltonian does not change under complex conjugation, we may write  $\corr{\psi|H |\psi}=-\corr{\psi| {\cal C} H {\cal C}|\psi}^* = -(\corr{\psi| {\cal C})^* H ({\cal C}|\psi})^*$. Now we notice that $({\cal C}|\psi\rangle)^*$ is equivalent to the MPS state $|\psi\rangle$ with $\phi_i\to-\phi_i$. Hence, when $\phi_i = 0$ we obtain $\corr{\psi|H |\psi}= -\corr{\psi|H |\psi} = 0$ for any values of $\theta_i$. This is sufficient to give a flow invariant subspace, $\dot \phi_i=0$ when $\phi_i=0$ according to Eq.~(\ref{eq:phidot}).

The above conditions of particle-hole symmetry and time reversal are met not only by the pure Hamiltonian of the PXP model, $H_{PXP}$, but also by any Hamiltonian which contains an odd number of $\sigma^x$ matrices. This motivates the choice of the deformation $H_{\mu_3}$ considered below and also in the main text for $K=3$ unit cell. In passing, we also note that all terms in the quasi-local deformation conjectured by Ref.~\cite{Soonwon2018} to give perfect quantum scars satisfy both of these symmetries, thus leaving the flow-invariant subspace intact.

Finally, we note that despite the angles $\phi_{i}$ being redundant when one restricts to the flow-invariant subspace, as we do in the next section, their presence is crucial for deriving the equations of motion (afterwards they can be set to zero). Indeed, if one sets the phases $\phi_i$ to zero before taking the derivatives in Eq.~(\ref{eq:thetadot}), one cannot obtain the equations of motion for amplitude variables.  One possible route to the equations of motion, if one wants to set the phase variables to zero from the very start, would be to directly minimize the quantum leakage (see Eq.~(\ref{Eq:leak}) below) with respect to $\theta_{1,2}$.

\subsubsection{EOMS for PXP with $K=3$ ansatz \label{App:Z3}}

In this section we derive the equations of motion for the three-site MPS ansatz with three variational parameters $(\theta_1,\theta_2,\theta_3)$. However, as we discussed above, such a derivation is more conveniently done with the help of conjugate variables,  $(\phi_1,\phi_2,\phi_3)$. Hence, in what follows we keep the phase variables and set them to zero at the end of the calculation. 

All calculations are done analogously to the previous case of $K=2$ site unit cell. Hence, we omit the details of the calculations, listing the resulting expressions. The right dominant vector, is calculated to be
\begin{eqnarray}\label{Eq:R-3unitcell} \nonumber
|R)
&=& r(1/r,-\cos\theta_{1}\sin\theta_{1},-\cos\theta_{1}\sin\theta_{1}, -\sin^2\theta_{1}), \\ \nonumber
&&\text{where}\quad r = \frac{\cos^2\theta_{2}\cos^2\theta_{3}+\sin^2\theta_{3}}{\cos^2\theta_{2}\sin^2\theta_{1}-1}.
\end{eqnarray}
Using the form of the right dominant vector, we calculate the normalization factor,
\begin{equation}
(L_\text{max}|R_\text{max}) = 1-\frac{\cos^2{\theta_{2}}\cos^2{\theta_{3}}+\sin^2{\theta_{3}}}{\cos^2{\theta_{2}}-1/\sin^2{\theta_{1}}}.
\end{equation}
The functions $f_i$ read:
\begin{equation}
f_{1} = -\frac{L}{3}\frac{ \sin ^2\theta_{2} \left(\cos ^2\theta_{3}+\sin ^2\theta_{1}\sin ^2\theta_{3}\right)}{1+\sin ^2\theta_{1}\sin ^2\theta_{2} \sin ^2\theta_{3}},\\
\end{equation}
\begin{equation}
f_{2} = \frac{L}{3}\frac{ \sin ^2\theta_{3} \left(\sin ^2\theta_{1} \cos ^2\theta_{2}-1\right)}{\sin ^2\theta_{1} \sin ^2\theta_{2} \sin ^2\theta_{3}+1},\\
\end{equation}
\begin{equation}
f_{3} = -\frac{L}{12}\frac{  3-2 \sin ^2\theta_{2} \cos (2 \theta_{3})+\cos (2 \theta_{2})}{ \sin ^2\theta_{2} \sin ^2\theta_{3}+\sin ^{-2}\theta_{1}}.
\end{equation}
\begin{widetext}
The local terms from the PXP Hamiltonian and deformation $ H_{\mu_3}$, Eq.~(\ref{Eq:h3}), contracted with the environment, result in the following matrices:
\begin{equation} 
\mathcal{H}^{i,i+1,i+2}_{PXP} = N_{PXP}\left(\begin{matrix} \cos ^2\theta_{i} \sin (2\theta_{i+2}) & 0 & 0 & 0 \\
 2 \sin \theta_{i} \cos \theta_{i} \sin \theta_{i+2} \cos\theta_{i+2} & 0 & 0 & 0 \\
 2 \sin \theta_{i} \cos \theta_{i} \sin \theta_{i+2} \cos \theta_{i+2} & 0 & 0 & 0 \\
 \sin ^2\theta_{i} \sin (2 \theta_{i+2}) & 0 & 0 & 0 \\ \end{matrix} \right), \\
\qquad
\mathcal{H}^{i,\ldots,i+5}_{\mu{3}}=N_{\mu_{3}}\left(\begin{matrix}  \cos ^2\theta_{i} & 0 & 0 & 0 \\
  \sin \theta_{i} \cos \theta_{i} & 0 & 0 & 0 \\
  \sin \theta_{i} \cos \theta_{i} & 0 & 0 & 0 \\
  \sin ^2\theta_{i} & 0 & 0 & 0 \\ \end{matrix} \right),
\end{equation}
where the indices $i$ are understood to be $\mathop{\rm mod} 3$ and we use the following short-hand notations:
\begin{equation} 
N_{PXP} = -(\cos{\theta_{i+1}}\sin{\phi_{i+1}})^{-1},\\ 
\quad
N_{\mu{3}} = -2\mu_3/(\sin\theta_{i}\sin\theta_{i+1}\sin\theta_{i+2}\sin\Omega\cos^2\theta_{i+1}),\\
\quad
\Omega= \phi_{i}+\phi_{i+1}-\phi_{i+2}.
\end{equation}
Substituting these matrices in an expression analogous to Eq.~(\ref{Eq:corrHPXP}), one can obtain the expectation value of the Hamiltonian. It has a cumbersome form and we do not need it here, since it vanishes when $\phi_i=0$.  Plugging $\corr{\psi|H_{PXP}+H_{\mu_3}|\psi}$ and expressions for $f_i$ into Eq.~(\ref{eq:thetadot}), we obtain the equations of motion for variables $\theta_i$. Since we are interested in the flow-invariant subspace, we set $\phi_i=0$ resulting in the following equations of motion:
\vspace{0.2 cm}
\begin{subequations}
\begin{multline}
M_1 \dot{\theta}_{1}=\sin \theta_{2} \big(\mu_{3} \cos \theta_{1} \sin \theta_{3} \big(6 \sin ^2\theta_{1}  \cos (2 \theta_{2})+3 \cos (2 \theta_{1})-19\big)+
4 \sin \theta_{2} \big(\mu_{3} \sin ^2\theta_{1} \cos\theta_{1} \sin\theta_{2}\sin (3 \theta_{3})+\cos (3 \theta_{3})\big)\big)+\\
   2 \sin\theta_{1} \sin (2 \theta_{2}) \big(-2 \cos ^2\theta_{1}\cos (2 \theta_{3})+\cos (2 \theta_{1})-3\big)-2 (3 \cos (2 \theta_{2})+5) \cos \theta_{3},
\end{multline}
\begin{multline}
 M_{2}  \dot{\theta}_{2}
 =16 \mu_{3} \sin ^3\theta_{1} \cos (3\theta_{2}) \sin ^3\theta_{3}+\mu_{3} \cos \theta_{2} \left(4 \sin ^3\theta_{1} \sin (3 \theta_{3})-73 \sin \theta_{1} \sin \theta_{3}+3 \sin (3 \theta_{1}) \sin \theta_{3}\right)\\-
   4
   \sin (2 \theta_{3}) \left((\cos (2 \theta_{1})+7) \sin\theta_{2}-2 \sin ^2\theta_{1} \sin (3 \theta_{2})\right)-8 \cos \theta_{1} \left(4 \cos ^2\theta_{1} \cos (2 \theta_{3})+5\right)+8 \cos (3 \theta_{1})
\end{multline}
\begin{multline}
M_{3} \dot{\theta}_{3}=  \mu_{3} \sin \theta_{1}\sin \theta_{2} \cos\theta_{3} \big(2 \sin ^2\theta_{1} \cos (2 \theta_{2})+\cos (2 \theta_{1})-17\big)-2 \sin ^2\theta_{2} \big(2 \mu_{3} \sin ^3\theta_{1} \sin \theta_{2} \cos (3\theta_{3})+\\
\sin (2 \theta_{1}) \sin (3 \theta_{3})\big)+\sin (2 \theta_{1}) (\cos (2 \theta_{2})+7) \sin \theta_{3}+8 \cos (2 \theta_{1}) \cos ^3\theta_{2}+10 \cos \theta_{2}-2 \cos (3 \theta_{2}),
\end{multline}
\end{subequations}
where the factors that multiply the derivatives read:
\begin{equation}
M_{1} = -4(3+\cos(2\theta_{2})-2\cos(2\theta_{3})\sin^2\theta_{3}),\ 
M_{2} = -64(\cos^2\theta_{3}+\sin^2\theta_{1}\sin^2\theta_{3}),\ 
M_{3} = 16(1-\cos^2\theta_{2}\sin^2\theta_{1}).
\end{equation}
\end{widetext}

\subsection{TDVP for TFIM \label{App:TFIM}}

We begin by motivating the form of the TDVP ansatz used in the Ising model, Eq.~(\ref{Eq:TFIM-ansatz}). Its form can be obtained by considering the approximate ``trotterized'' unitary evolution,
\begin{equation}\label{Eq:unit}
e^{-iH_\text{TFIM} \delta t} \approx \prod_{i}e^{-i(J \sigma_{i}^{z}\sigma_{i+1}^{z}+h_{z}\sigma^{z}_{i})\delta t} \prod_{i}e^{-i h_{x}\sigma^{x}_{i}\delta t},
\end{equation}
where the approximation works for small values of $\delta t$. The operator on the right-hand side can be written as a translationally-invariant matrix product operator (MPO)~\cite{Pirvu_2010} with bond-dimension $\chi=2$. This MPO has two physical indices, $\sigma_1, \sigma_2 = \uparrow,\downarrow$ and can be written as:
\begin{equation}
\begin{split}
&M^{\uparrow \uparrow} = \left(\begin{matrix}\cos \tilde\theta e^{-i(\tilde\chi + \tilde\phi)} & \cos \tilde\theta e^{-i(-\tilde\chi + \tilde\phi)}\\
0&0 \end{matrix}\right)\\
&M^{\uparrow \downarrow} = \left(\begin{matrix}-i\sin \tilde\theta e^{-i(\tilde\chi + \tilde\phi)} & -i\sin \tilde\theta e^{-i(-\tilde\chi + \tilde\phi)}\\
0&0 \end{matrix}\right)\\
&M^{\downarrow \uparrow} = \left(\begin{matrix} 0 & 0 \\ -i\sin \tilde\theta e^{i(\tilde\chi + \tilde\phi)} & -i\sin \tilde\theta e^{i(-\tilde\chi + \tilde\phi)}\end{matrix}\right)\\
&M^{\downarrow \downarrow} = \left(\begin{matrix} 0 & 0 \\ \cos \tilde\theta e^{i(\tilde\chi + \tilde\phi)} & \cos \tilde\theta e^{i(-\tilde\chi + \tilde\phi)}\end{matrix}\right).
\end{split}
\end{equation}
Values of the angles are found to be $(\tilde\theta,\tilde\phi,\tilde\chi) = (h_{x}\delta t,h_{z}\delta t,J\delta t)$. 

We can obtain an MPS from this MPO by applying it to a reference state. For convenience, we choose the reference state to be $\ket{\uparrow\uparrow\ldots}$ state. This results in the MPS with $\chi=2$ that is specified by matrices:
\begin{equation}\label{Eq:TFIM-guess}
\begin{split}
A^{\uparrow} = M^{\uparrow \uparrow}, \quad A^{\downarrow} = M^{\downarrow \uparrow}.
\end{split}
\end{equation}
Now we allow the angles to be independent parameters, obtaining an MPS ansatz with three variational parameters. Two of these parameters are phase variables $(\tilde\phi,\tilde\chi)$ and one is an amplitude variable $\tilde\theta$. Hence, it is natural to extend this ansatz by adding yet another amplitude variable $\xi$ that is conjugate to $\tilde\chi$. This can be done by replacing $e^{-i \tilde\chi}\rightarrow \cos{\tilde\xi}e^{-i \tilde\chi}$ and $e^{i \tilde\chi}\rightarrow \sin{\tilde\xi}e^{i \tilde\chi}$.  After such an extension, supplemented by a multiplication with the phase $e^{i\tilde\phi}$ and a shift in parameters
\begin{equation}
\chi= -2 \tilde{\chi}, \quad \phi = 2\tilde{\phi}-\frac{\pi}{2},
\end{equation}
Eq.~(\ref{Eq:TFIM-guess}) turns into the MPS ansatz~(\ref{Eq:TFIM-ansatz}) used in the main text.

The MPS ansatz~(\ref{Eq:TFIM-ansatz}) was obtained as the twist of a product state by a $\chi=2$ MPO operator that approximates the unitary evolution with the Hamiltonian of transverse-field Ising model. Thus, this state can be understood as the generalization of the product-state mean-field ansatz that incorporates short-range entanglement generated by unitary evolution with $H_\text{TFIM}$. Indeed, one can explicitly check that by restricting the angles to values $(\chi,\xi)=(0,\pi/4)$ we obtain a generic translationally invariant product state parametrized by $(\theta,\phi)$.

After justifying the form of the ansatz, we discuss its equations of motion. The unit cell consists of $K=1$ site, thus the transfer matrix can be obtained straightforwardly. Its largest eigenvalue reads
\begin{equation}
\lambda_\text{max} = \frac{1}{8}(2+2 \cos{2\xi}+\sqrt{6+2\cos{4\xi}+8\cos{2\xi}\cos{4\theta}}).
\end{equation}
This eigenvalue is doubly degenerate when the expression under the square root vanishes. In what follows we ignore the degenerate surface and assume that the density matrix is non-degenerate. Crucially, the largest eigenvalue is generally different from one, $\lambda_\text{max}\neq 1$. Hence, this ansatz does not result in the normalized wave function and we have to modify the derivation of our equations of motion. 

Even though  TDVP can be reformulated to apply to non-normalized states~\cite{kramer1981geometry}, we prefer to normalize each tensor in the numerical implementations $\tilde{A}\rightarrow A/\sqrt{\lambda_\text{max}}$, so that $\tilde{\lambda}_\text{max} = 1$. The rest of the calculation is similar to the previous section. The main difference is that due to the increased complexity of the tensors and the presence of normalization in tensors,  analytical calculations are more complicated even if the method is exactly the same as in the PXP model. The simplest way to avoid the complexity is to evaluate all expressions including the derivatives numerically, however this may result in instabilities in the integration of the EOMS. In our implementation we evaluate all derivatives analytically but substitute numeric values for the variables for which no more derivatives are to be taken. We find that using such approach is computationally efficient, while providing numerically stable time integration. Resulting dynamical system can be found in Ref.~\SOM.
\begin{figure*}[t!]
\begin{center}
\includegraphics[width=1.99\columnwidth]{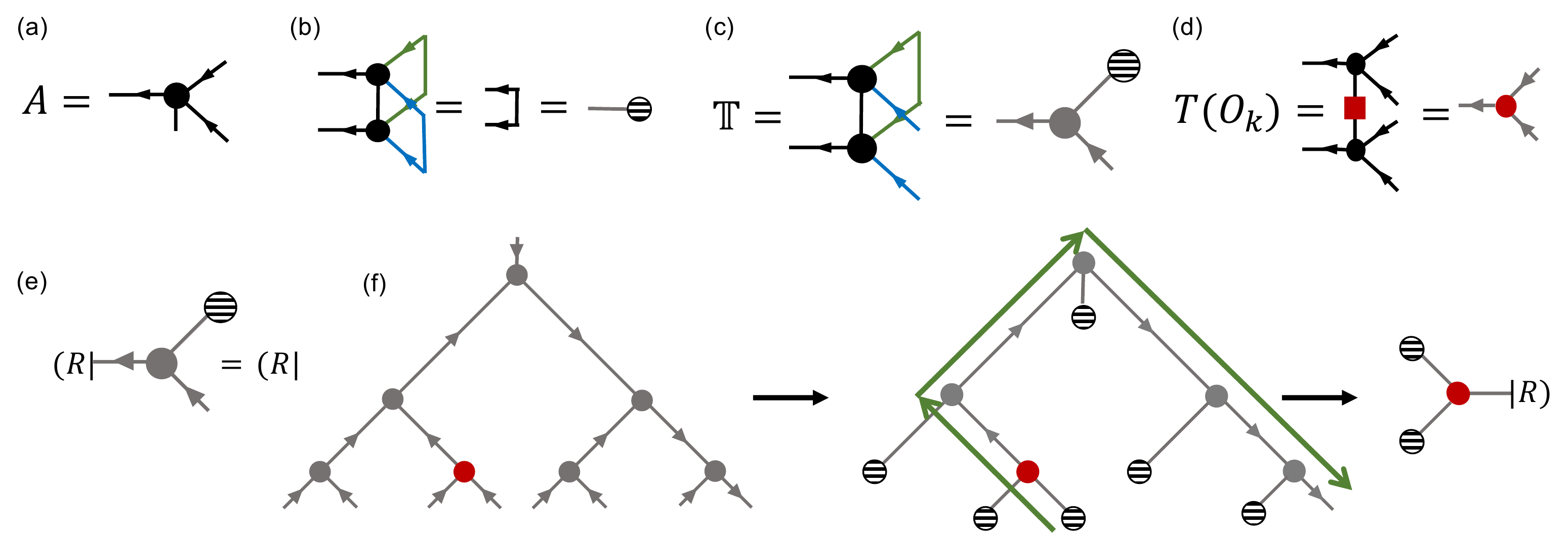}
\caption{ \label{Fig:Tree}
Illustration of the calculation of the one-point correlation function for $C=3$. The gray colored bonds denote a vectorization of the bonds pointing towards the same direction when the contractions are performed. Graphical definitions of: (a) The local tensor $A$.  (b)  The gauge symmetry, Eq.~(\ref{Eq:gauge}), where hatched circle is introduced as a short-hand notation for a Kronecker delta in virtual index space. (c) The local transfer matrix, Eq.~(\ref{Eq:GenTM}). (d) The contraction of the local operator $O_{k}$ by the local tensor and (e) the right dominant eigenvalue of the transfer matrix. Panel (f) illustrates the calculation of Eq.~(\ref{Eq:corfunc_tree}) in three steps. In the first step we show a part of the tree which includes the site $k$, where the local operator acts. The gray colored vertices imply contracted tensors in those sites.  In the second step the gauge symmetry (b) is used and the whole tree is reduced to a semi-infinite one dimensional tensor network (green line is a guide to the eye). In the final step the infinite product of local transfer matrices is replaced by the right dominant eigenvector of the matrix, see Eq.~(\ref{Eq:corfunc_tree}).
}
\end{center}
\end{figure*} 
\section{TDVP for tree tensor states \label{App:tree}}

In this section, we introduce the TTS ansatz for trees of arbitrary connectivity. Afterwards we discuss how to reduce the calculation of local correlation functions of a TTS to a calculation for an effective one dimensional tensor network. This allows to use the method described in Section~(\ref{App:Z2}) for calculating equations of motion. In Fig.~(\ref{Fig:Tree}) we give a graphical illustration of the algebraic calculations presented in this section. Finally, we discuss the relation between the TTS ansatz and the projected entangled pair states (PEPS) ansatz~\cite{verstraete2004renormalization}.

\subsection{Tensor tree state ansatz \label{App:TTS}}

The basic building block of the TTS ansatz is the tensor $A^{\sigma^z}_{i_{1,\ldots, i_C}}$ with $C$ virtual indices $i_{1,\ldots C}$ and one physical index, $\sigma^z$ see Fig.~\ref{Fig:Tree}(a). Out of $C=3$ virtual indices shown in Fig.~\ref{Fig:Tree}(a) one is designated as ``output leg'' with an outgoing arrow; the other indices are input legs.  In order to obtain the normalized wave function that obeys the Rydberg blockade constraint, one may choose the following form of tensors:
\begin{equation}\label{Eq:TTS_full}
\begin{split}
A^{\downarrow}_{i_{1},\dots,i_{C-1},0} &= \cos^{I}{\theta}\sin^{C-1-I}{\theta},\\
A^{\uparrow}_{0,\ldots 0,1} &= i e^{-i \phi},\\
\end{split}
\end{equation}
with all other elements of $A$ being zero. These expressions generalize Eq.~(\ref{Eq:TTS}) for arbitrary connectivity, $C$. Such parametrization produces an ``output'' 1-bit, $i_C=1$ when the corresponding site is excited, and 0-bit, $i_C=0$ when the given site is in the ground state, $\sigma^z=\downarrow$. In turn, the quantity  $I = \sum^{C-1}_{j=1} i_{j}$ in Eq.~(\ref{Eq:TTS_full}) counts the number of ``excitations''  on $C-1$ adjacent sites. Only in the case when all of these $C-1$ adjacent sites do not have an excitation, $i_1=\ldots=i_{C-1}=0$, can the given site  be in the excited state, thus implementing the constraint of no adjacent $\uparrow$ states.  

We selected out one of the virtual indices by designating it as ``output''. However, the particular choice of this leg is not important, and the TTS ansatz of Eq.~(\ref{Eq:TTS_full}) is invariant under permutations of all remaining input legs,
\begin{equation}\label{Eq:Psym}
A_{i_{1},\ldots,i_{C-1},i_{C}} = A_{\mathcal{P}(i_{1},\ldots,i_{C-1}),i_{C}},
\end{equation}
where $\mathcal{P}\in S_{C-1}$ and $S_{C-1}$ denotes the symmetric group of $C-1$ symbols. This property  allows us to swap input legs at will, and it also holds for TTS with two alternating tensors that we use to imitate states with $K=2$ sites unit cell on a lattice. 

As we discussed, the  TTS ansatz with tensors in Eq.~(\ref{Eq:TTS_full}) respects the Rydberg blockade constraints. This constraint imposes restrictions on the non-zero elements of the tensor $A$. The particular form of these elements in Eq.~(\ref{Eq:TTS_full}) is dictated by the normalization condition. For the bipartite tree, one can check that this ansatz generates a wave function 
\begin{equation}\label{Eq:variational}
\ket{\psi} = N\sum c^{L/2-n_1}_{\downarrow_{1}} c^{n_1}_{\uparrow_{1}} c^{L/2-n_2}_{\downarrow_{2}} c^{n_2}_{\uparrow_{2}}\ket{\{n_1,\uparrow_1\}\{n_2,\uparrow_2\}}
\end{equation}
where the sum runs over all product states that obey the constraint, and $N$ ensures the normalization in the thermodynamic limit and is calculated in Eq.~(\ref{Eq:normfactor}). Numbers $n_{1,2}$ count the number of $\uparrow$ states in the corresponding sublattice, which are assigned the following weights: 
\begin{subequations}\label{Eq:weights}
\begin{eqnarray}
c_{\downarrow_{1}} &=& \cos{\theta_{1}},\quad c_{\uparrow_{1}} =ie^{-i \phi_{1}}\tan{\theta_{2}},\\
c_{\downarrow_{2}}&=& \cos{\theta_{2}},\quad c_{\uparrow_{2}} =ie^{-i \phi_{2}}\tan{\theta_{1}}.
\end{eqnarray}
\end{subequations}
Note that the weights are the same for all connectivities. This property is reminiscent of the mean field product state ansatz where the weights would be $c_{\downarrow_{i}} =\cos{\theta_{i}}$, $c_{\uparrow_{i}} = -ie^{-i \phi_{i}}\sin{\theta_{i}}$. However, in the presence of a constraint these mean-field weights would lead to a wave function with the norm vanishing  as $||\ket{\psi}||\propto e^{-a L}$ in the thermodynamic limit.

In addition to being normalized, one can check that the tensor  $A$ obeys a generalized canonical gauge, see Fig.~\ref{Fig:Tree}(b),
\begin{equation}\label{Eq:gauge}
\sum_{i_{1},\ldots i_{C-1},\sigma^z}A^{\sigma^z}_{i_{1},\ldots i_{C-1},i_{C}}(A^{\sigma^z}_{i_{1},\ldots i_{C-1},i_{C}'})^* = \delta_{i_{C},i_{C}'},
\end{equation}
for arbitrary values of $\theta,\phi$. This condition is enough to allow for normalization in thermodynamic limit. Moreover, it greatly simplifies the calculation of expectation values as we discuss below.

\subsection{Mapping the tensor tree to a one-dimensional lattice \label{App:TTS1D}}

As explained in Section~\ref{TDVP_formalism}, the main ingredient of TDVP are one-point correlation functions of the form $\braket{\psi|\partial_{x_k}|\psi}$ (or similar expressions in which the local operator is the Hamiltonian density), where $k$ denotes the site of the lattice. For clarity we work with a one site unit cell $K=1$, however the results also hold for $K=2$. We define the generalized transfer matrix, see Fig.~\ref{Fig:Tree}(c), as:
\begin{equation}\label{Eq:GenTM}
\mathbb{T}_{(i_{1}i'_{1}),(i_{C} i_{C}')} = \sum_{i_{2},\ldots i_{C-1},n}A^{n}_{i_{1},\ldots i_{C-1},i_{C}}(A^{n}_{i'_{1},i_{2},\ldots i_{C-1},i_{C}'})^* ,
\end{equation}
where the indices inside parenthesis are vectorized. For $C=2$, $\mathbb{T}$ is the same as the single site transfer matrix defined in Eq.~(\ref{Eq:transfer}). We note that tracing any set of $C-2$ input indices is equivalent due to the permutation symmetry of the input legs, Eq.~(\ref{Eq:Psym}). Due to the canonical gauge of Eq.~(\ref{Eq:gauge}), the left eigenvector of the transfer matrix is just the identity, ${(L|_{a,b} = \delta_{a,b}}$. 
We also define the generalized transfer matrix that includes local operator [see Fig.~\ref{Fig:Tree}(d)], 
\begin{equation}
T(O_{k})=  \sum_{n',n}O_{n',n}A^{n}_{i_{1},\ldots i_{C-1},i_{C}}(A^{n'}_{i'_{1},i'_{2},\ldots i'_{C-1},i_{C}'})^*.
\end{equation}
Using the canonical gauge [Fig.~\ref{Fig:Tree}(b)] and property of the right eigenvector [Fig.~\ref{Fig:Tree}(e)], we write the one-point correlation functions in Fig.~\ref{Fig:Tree}(f) as:
\begin{eqnarray}\label{Eq:corfunc_tree}
\braket{\psi|O_{k}|\psi}&=&\lim_{q\rightarrow \infty}(L_1 L_2 \dots L_{C-1}|T(O_{i}) \mathbb{T}^{q}_{C}|R)\\\nonumber
&=&(L_1 L_2 \dots L_{C-1}|T(O_{i})|R),
\end{eqnarray}
where $(L_l L_m| \equiv (L_l| \otimes (L_m|$. The subscript indicates which input bonds are traced by the corresponding vector, while the transfer matrix $\mathbb{T}_{C}$ is applied to the output bonds ($i_{C},i_{C}'$), see Fig.~\ref{Fig:Tree}(d). 
Equation~(\ref{Eq:corfunc_tree}) has the remarkable feature of being able to trace the whole (infinite) lattice by the repeated application of a transfer matrix of system size independent dimensions. Such a feature is reminiscent of the one-point function calculation using MPS, see Eq.~(\ref{Eq:corfunc_1d}), and reflects the loop-free structure of TTS. The calculation of one-point functions of local operators which are supported in more than a single site is a straightforward generalization of Eq.~(\ref{Eq:corfunc_tree}).

The normalization factor $N$, of  Eq.~(\ref{Eq:variational}) is derived by calculating $\braket{\psi|\psi}$, which is a trivial modification of Eq.~(\ref{Eq:corfunc_tree}) and corresponds to
\begin{equation}\label{Eq:normfactor}
N = (L|R)^{-1/2} =( 1 + \cos^2\theta_{1}\tan^2\theta_{2})^{-\frac{1}{2}}.
\end{equation}

\subsection{Comparing variational wave function on a lattice and TTS\label{App:TTS_PEPS}}

In the previous section we explained the main steps in the analytical calculation of the TDVP equations of motion on a TTS. However, our goal is to approximate the dynamics on higher dimensional lattices of the same connectivity but different topology, e.g. square lattice. The true variational wavefunction for the square lattice is also the form of Eq.~(\ref{Eq:variational}), but with a different normalization factor. Such a wavefunction is a tensor network which has the same geometry as the underlying lattice, so for the lattices of interest it belongs to the PEPS family~\cite{verstraete2004renormalization}. 

The calculation of one point correlation functions, $\braket{\psi |O_{k}|\psi}$ in PEPS is known to be a sharp P-Hard problem for generic PEPS states \cite{PhysRevLett.98.140506} and is associated with the presence of loops in the lattice. To circumvent this problem, numerical algorithms typically employ some form of truncation in the calculation of correlation functions. However, the analytical calculation of the equations of motion is practically impossible for an infinite PEPS. In the main text we mention the comparison of the the dynamics on the true variational wavefunction for the lattice (PEPS) and the TTS. We performed the comparison by parametrizing the PEPS ansatz as Eq.~(\ref{Eq:variational}). We use a finite system with periodic boundaries (4 by 4 lattice), in order to be able to calculate the correlation functions analytically. The TDVP equations of motion are calculated for the unnormalized state (not including the factor $N$), to avoid dealing with the complicated normalization factor.

\section{Classifying trajectories by quantum leakage\label{App:leak}}

In this Appendix we provide additional details on how to calculate the rate at which the quantum system is leaving the MPS variational manifold -- which we call quantum leakage. We start with the general framework and provide detailed justification of the fidelity bound. Afterwards, we apply quantum leakage to characterize the different periodic trajectories observed in the deformed PXP model. 

\subsection{General framework: quantum leakage }

If one initializes the quantum state in the form of Eq.~(\ref{Eq:psi-A}), at $t=0$ the values of all local observables and their time derivative are captured by TDVP equations of motion. At later times the TDVP evolution begins to disagree with the exact unitary dynamics, see Fig.~\ref{Fig:sketch}. Intuitively, this is visualized by  ``leakage'' of the exact quantum wave function from the variational manifold. The instantaneous leakage rate is given by the disagreement between quantum evolution and TDVP dynamics,
\begin{multline}\label{Eq:leak}
\Lambda^2(\{\vec x_i\}) = ||\ket{\dot{\psi}}+ iH\ket{\psi} ||^2 =\braket{\psi|H^2|\psi}
\\  
-2\sum_{i}\dot{{x}}_{a}\text{Im}(\braket{\partial_{{x}_{a}}\psi|H|\psi}) 
+\sum_{a,b}\dot{{x}}_{a}\text{Re}(\braket{\partial_{{x}_{a}}\psi|\partial_{{x}_{b}}\psi})\dot{{x}}_{b}.
\end{multline}
We note that this quantity goes beyond TDVP equations of motion, as it contains the square of the Hamiltonian operator, $H^2$, which depends on quantum commutation rules and operator algebra. While we were able to obtain the equations of motion, Eq.~(\ref{Eq:TDVPMPS}), without explicit calculation of two-body correlators, this is not possible in the present case. Nevertheless, the use of the gauge of Eq.~(\ref{Eq:phase}) allows to cancel all disconnected correlators. This can be accounted by replacing $\corr{*} \to \corr{*}_c$ in Eq.~(\ref{Eq:leak}), where the connected correlators are defined as
\begin{eqnarray}\label{Eq:connectedleak}
\braket{\partial_{\textbf{x}_{i}}\psi|\partial_{\textbf{x}_{j}}\psi}_{c}
&=& \braket{\partial_{\textbf{x}_{i}}\psi|\partial_{\textbf{x}_{j}}\psi} -  \braket{\partial_{\textbf{x}_{i}}\psi|\psi}\braket{\psi|\partial_{\textbf{x}_{j}}\psi},\\ \label{Eq:connectedleak2}
\braket{\partial_{\textbf{x}_{i}}\psi|H|\psi}_{c}
&=&\braket{\partial_{\textbf{x}_{i}}\psi|H|\psi}-\braket{\partial_{\textbf{x}_{i}}\psi|\psi}\braket{\psi|H|\psi},\\ \label{Eq:connectedleak3}
\braket{\psi|H^2|\psi}_{c}&=& \braket{\psi|H^2|\psi}-\braket{\psi|H|\psi}^2.
\end{eqnarray}
Upon substitution of the connected correlators the error scales as $\Lambda^2\propto L$. As we show in the next section, the disconnected component that is proportional to $L^2$ vanishes. In the thermodynamic limit the quantity of interest is the error density, $\gamma^2 = \Lambda^2/L$, as shown below.

In contrast to EOMS, the error calculation is more complicated because it involves two-point correlators. We briefly describe the method presented in~\cite{Haegeman} on how to resum two-point correlators in the thermodynamic limit and present an analytical recipe used to calculate the leakage in this work. We note that a similar calculation was performed in Ref.~\cite{wenwei18} for the flow invariant subspace (${\phi}_i =0$, $\braket{H} = 0$) of the 2-site PXP model. 

We are interested in calculating the correlators defined in Eq.~(\ref{Eq:connectedleak}). We show how to perform the resummation of two-point correlators of a uniform MPS, initially displayed in Ref.~\cite{Haegeman}. Extensions to larger unit cells follow straightforwardly by using the transfer matrix of the corresponding unit cell. For two parameters $a,b$ we get,
\begin{multline}\label{Eq:GramMat}
\braket{\partial_{b}\psi|\partial_{a}\psi} =  \sum_{m,n} \\
 \raisebox{-0.35in}{\includegraphics[width=0.7\columnwidth]{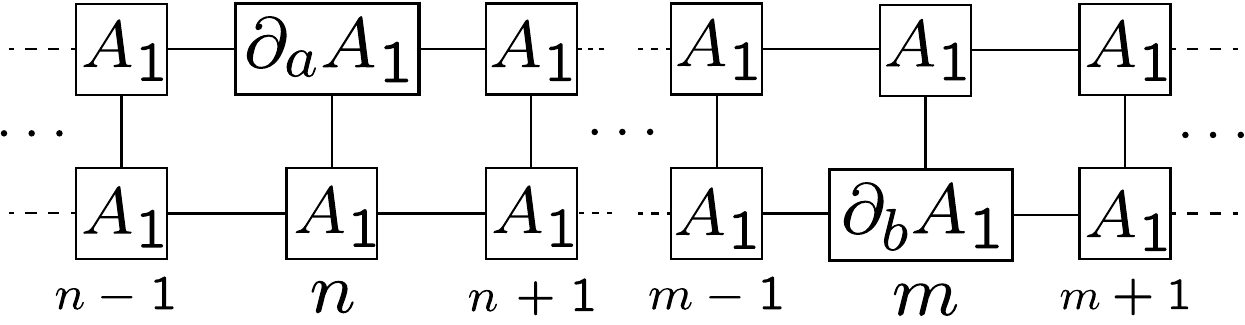}}=
 \\
L \frac{(L|T^{\partial_{a}A}_{\partial_{b}A}|R)}{(L|R)} 
+
\sum^{L-1}_{n=1}\sum_{q=0}^{L-n-1} \frac{(L|T^{\partial_{a}A}[T_{u.c.}]^{q} T_{\partial_{b}A}|R)}{(L|R)}+
\\
\sum^{L-1}_{n=1}\sum_{q=0}^{L-n-1} \frac{(L|T_{\partial_{b}A}[T_{u.c.}]^{q} T^{\partial_{a}A}|R)}{(L|R)},
\end{multline}
where $n,m$ are lattice site labels and in the second line we replaced the summation over $n,m$ by a sum over $n$ and $q=|n-m-1|$. From such expressions it is evident that one has to resum geometric series of the form $\sum_{q}[T_{u.c.}]^{q}$, which is possible if the operator has spectral radius $\rho(T_{u.c.})<1$. Since the transfer matrix has a single largest unit eigenvalue, the dominant subspace has to be projected out and resummed separately,
\begin{multline}\label{Eq:resum}
\sum_{q=0} [T_{u.c.}]^{q} = \sum_{q=0} \left(\mathcal{Q}T_{u.c.}\mathcal{Q} + \mathcal{P}\right)^q 
= \sum_{q=0} \mathcal{P}+\\
 \mathcal{Q}\left(\mathds{1}+\sum_{q=1} [\mathcal{Q}T_{u.c.}\mathcal{Q}]^{q}\right)\mathcal{Q}
 = \mathcal{T}^{-1} + \sum_{q=0} \mathcal{P},
\end{multline}
where we have defined the projector onto the dominant subspace, $\mathcal{P} = |R)(L|/(L|R)$,  its complement, $\mathcal{Q}=\mathds{1}-\mathcal{P}$, and introduced the matrix $\cal T$ as 
\begin{equation}
 \mathcal{T}^{-1}= \mathcal{Q}(\mathds{1} - \mathcal{Q}T_{u.c.}\mathcal{Q})^{-1}\mathcal{Q}.
\end{equation}
Note that the same resummation formula holds for any size of the unit cell.
Substituting Eq.~(\ref{Eq:resum}) into Eq.~(\ref{Eq:GramMat}) we get,
\begin{multline}\label{Eq:GramMatfull}
\braket{\partial_{b}\psi |\partial_{a}\psi}=
L(L-1) \frac{(L|T_{\partial_{b} A}\mathcal{P}T^{\partial_{a} A}|R)}{(L|R)}
\\
+L\frac{
(L|T^{\partial_{a} A}_{\partial_{b} A}+T_{\partial_{b} A}\mathcal{T}^{-1} T^{\partial_{a} A}+T^{\partial_{a} A}\mathcal{T}^{-1} T_{\partial_{b} A}|R)
}
{(L|R)}.
\end{multline}
The first line contains the resummation of disconnected correlator, and these terms are cancelled by the proper gauge choice. The terms in the second line correspond to cases when both derivatives are taken in the same unit cell and long range resummations of correlations for cases when $n>m$ and $m<n$.  

The expressions for $\braket{\psi|H^2|\psi}$ and $\braket{\partial_{b}\psi|H|\psi}$  are also calculated using   Eq.~(\ref{Eq:resum}). For the Hamiltonian which can be written as a sum of local operators $H= \sum_{i}h_{i}$, with $h_{i}$ having support on an finite number of sites, chosen to be two for simplicity, we find: 
\begin{multline}
\label{Eq:Hamder}
\braket{\partial_{b}\psi |H|\psi}=L(L-2)\frac{(L|\mathcal{H}\mathcal{P}T_{\partial_{b} A}|R)} {(L|R)}
\\
+
L\frac{(L|\mathcal{H}_{\partial^k_{b}A}+ \mathcal{H}\mathcal{T}^{-1} T_{\partial_{b} A}+T_{\partial_{b} A}\mathcal{T}^{-1} \mathcal{H} |R)}{(L|R)},
\end{multline}
\begin{multline}\label{Eq:Hamsq}
\braket{\psi |H^2|\psi}=L\left(L-5\right)\frac{(L|\mathcal{H}\mathcal{P}\mathcal{H}|R)}{(L|R)}
\\
+L\frac{(L|\mathcal{H}^{(2)} +2\mathcal{H}\mathcal{T}^{-1} \mathcal{H}|R) }{(L|R)},
\end{multline}
where $\mathcal{H}$ is the contraction of the local Hamiltonian density with environment, and the matrices $\mathcal{H}_{\partial_{b}A}$, $\mathcal{H}^{(2)}$ are defined as
\begin{eqnarray}
{\cal H}_{\partial_a A} &=& \ \raisebox{-0.25in}{\includegraphics[width=0.4\columnwidth]{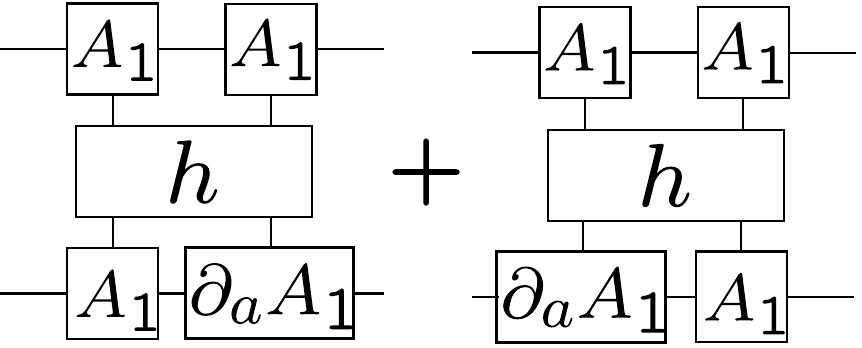}},\\
{\cal H}^{(2)} &=&\raisebox{-0.35in}{\includegraphics[width=0.17\columnwidth]{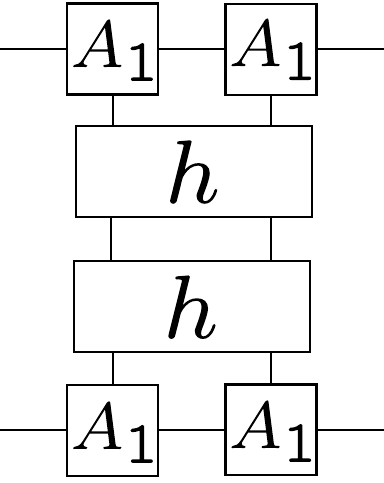}} +\raisebox{-0.35in}{\includegraphics[width=0.5\columnwidth]{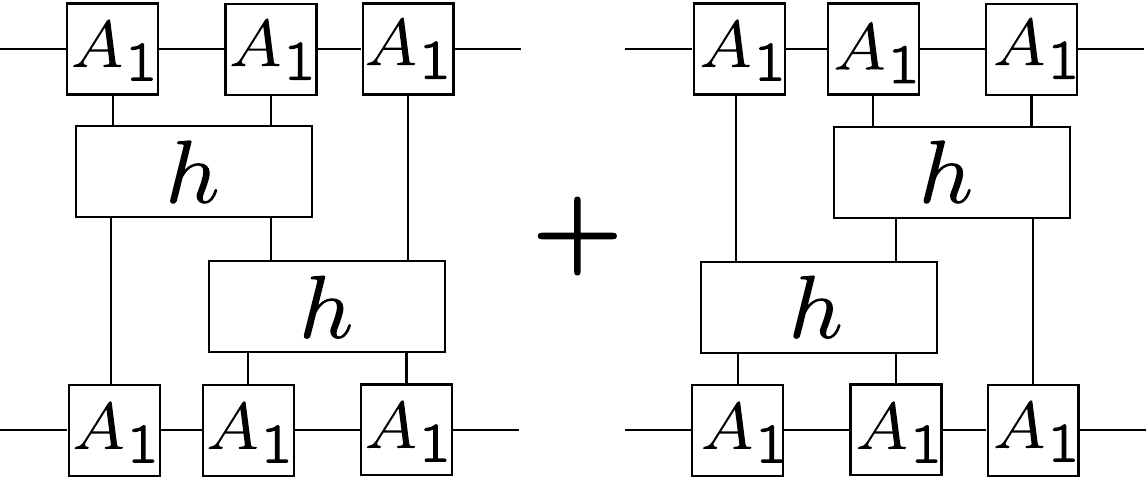}}.
\end{eqnarray}
These definitions are valid under the assumption that the Hamiltonian has only two-site terms, but generalization to longer-range terms in the Hamiltonian are straightforward. 

The disconnected correlators in Eqs.~(\ref{Eq:connectedleak})-(\ref{Eq:connectedleak3}) are calculated as:
\begin{eqnarray}
\braket{\partial_{b}\psi |\psi}\braket{\psi |\partial_{a}\psi} &=& \frac{L^2}{(L|R)}(L|T_{\partial_{b} A}\mathcal{P}T^{\partial_{a} A}|R),\\
\braket{\psi|H|\psi}\braket{\partial_{b}\psi |\psi} &=&\frac{L^2}{(L|R)}(L|\mathcal{H}\mathcal{P}T_{\partial_{b} A}|R),\\
\braket{\psi|H|\psi}^2&=& \frac{L^2}{(L|R)}(L|\mathcal{H}\mathcal{P}\mathcal{H}|R).
\end{eqnarray} 
Collecting together these expressions with Eqs.~(\ref{Eq:GramMatfull})-(\ref{Eq:Hamsq}) we observe that  terms in the connected correlators that scale as $L^2$ vanish. Thus, the instantaneous error scales linearly with the system size,  $\Lambda^2\propto L$.

\subsection{Fidelity bound}

\begin{figure}[t]
\begin{center}
\includegraphics[width=0.95\columnwidth]{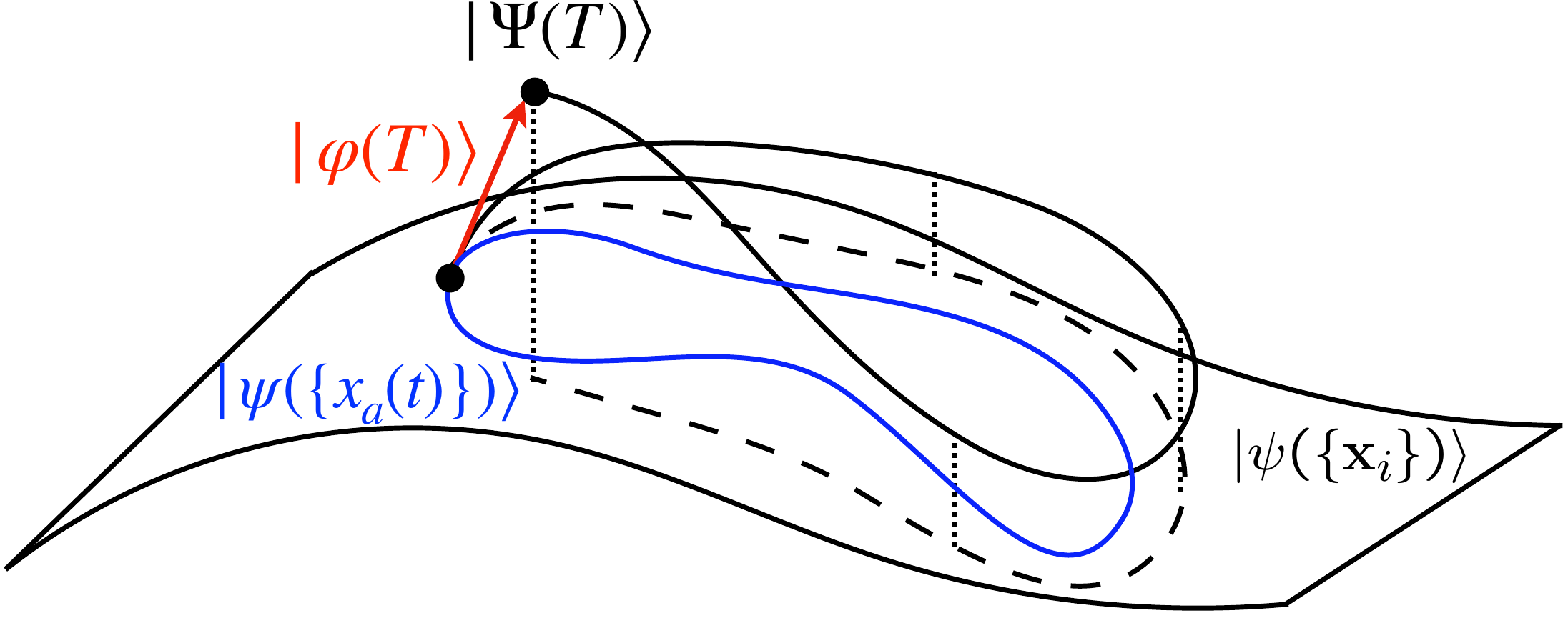}\\
\caption{ \label{Fig:sketch}
We derive the general bound on the norm of the vector $|\varphi(t)\rangle$ that represents the mismatch between the exact wave function and its TDVP ``image'' after the TDVP evolution returns to its initial state.  We note that the projection of the exact evolution onto TDVP manifold does not coincide with the TDVP trajectory at finite times.}
\end{center}
\end{figure}

At long times, a generic quantum many-body system is eventually expected to develop entanglement that cannot be captured by the MPS ansatz in Eq.~(\ref{Eq:psi-A}). At this point, the TDVP dynamics $\ket{\psi(\{x_a(t)\})}$ and exact unitary evolution, $e^{-iHt} \ket{\psi(\{x_a(0)\})}$, are expected to strongly disagree. However, as we demonstrate in the main text, the TDVP equations of motion are generically expected to have short periodic trajectories, see the blue line in Fig.~\ref{Fig:sketch}. If the quantum system were following the TDVP dynamics exactly, that would imply persistent oscillations in local observables and revivals of the many-body quantum fidelity to unity. Due to non-zero leakage, the quantum and TDVP evolution would disagree after a single period. In what follows we obtain a \emph{lower bound} on the many-body fidelity, Eq.~(\ref{eq:fid}), after the time $T$, i.e., the period of the trajectory.

To obtain the bound we represent the exact wave function, $\ket{\Psi}$, that follows the exact evolution according to the Schr\"odinger equation, $i\partial_t\ket{\Psi(t)} = H\ket{\Psi(t)}$, as a sum of two terms, 
\begin{equation}
  \ket{\Psi(t)} = \ket{\psi(\{x_a(t)\})} + \ket{\varphi(t)},
\end{equation}
where $\ket{\psi(\{x_a(t)\})}$ belongs to the variational manifold and is obtained from the TDVP equations of motion, while $\ket{\varphi(t)}$ is the error vector defined as the difference between these two wave functions, see Fig.~\ref{Fig:sketch}.

Expressing the error vector via $\ket{\Psi(t)}$ and the TDVP wave function, we obtain the following equation of motion:
\begin{multline}\label{Eq:dvar}
\partial_t \ket{\varphi(t)} = -iH \ket{\Psi(t)} -\ket{\dot \psi(\{x_a(t)\})} 
\\
=-iH \left(\ket{\psi(\{x_a(t)\})} + \ket{\varphi(t)}\right) -  \dot x_b(t) \ket{\partial_b\psi(\{x_a(t)\})}
  \\
   =- iH\ket{\varphi(t)} - \left[\ket{\dot \psi(\{x_a(t)\})}+i H \ket{ \psi(\{x_a(t)\})}\right],
\end{multline}
where we explicitly used the Schr\"odinger equation for~$\ket{\Psi(t)}$. The first term in the last line transports the error forwards in time and does not change the norm of $\ket{\varphi(t)}$. The second term describes the change of the error due to mismatch between exact evolution $\ket{\Psi(t)}$ and its TDVP projection, $\ket{\psi(\{x_a(t)\})}$, and its norm coincides with the quantum leakage defined in Eq.~(\ref{Eq:leak}).
Calculating the change in the norm of vector $\ket{\varphi}$ from Eq.~(\ref{Eq:dvar}) we obtain
\begin{equation}
\partial_t \langle \varphi | \varphi \rangle = - \bra{\varphi} \left(\partial_t+iH\right)\ket{ \psi(\{x_a(t)\})} + \text{h.c.},
\end{equation}
where we omit time dependence of $\varphi$ to simplify notations. Using the triangle and Cauchy-Schwartz inequalities we bound the growth of the norm of the error vector as
\begin{align} \nonumber
  \left|\partial_t \langle \varphi | \varphi \rangle \right|
  &\le 2 \left| \bra{\varphi}\left(\partial_t+iH\right)\ket{\psi(\{x_a(t)\})} \right| \\
  &\le 2 \| \varphi \| \, \|\left(\partial_t+iH\right) \ket{\psi(\{x_a(t)\})}\|,
\end{align}
where $\| \varphi \| = \sqrt{\langle \varphi | \varphi \rangle}$ is the norm of the error vector. Using this notation, we obtain an \emph{upper bound} on the rate of increase of the norm of $\ket{\varphi}$,
\begin{equation}
  \left|\partial_t \| \varphi \| \right| \le \Lambda(t) = \|\left(\partial_t+iH \right) \ket{\psi(\{x_a(t)\})} \|,
\end{equation}
where the quantity $\Lambda(t)$ is the instantaneous geometric error that was already defined in Eq.~(\ref{Eq:leak}).
Now, the triangle inequality can be used to bound the norm of the error vector accumulated during evolution over the finite period of time $t$, 
\begin{equation}
  \| \varphi(t) \| \le \int_0^t \mathrm{d}\tau\, \Lambda(\tau) = I_t
\end{equation}
by the integrated geometric error $I_t$.

The Fubini-Study metric or quantum angle can be defined as
\begin{equation}
  \gamma(a,b) = \arccos\frac{|\langle a | b \rangle|}{\|a\|\,\|b\|}\text{.}
\end{equation}
It is a natural metric on projective space, where pure states are represented by lines through the origin of Hilbert space, reflecting the fact that global phases are not physically meaningful. It is a quotient of the standard Euclidean metric restricted to the unit sphere. The distance $\gamma$ can be visualized as an angle between the two lines.

The three vectors $\ket{\Psi(t)}$, $\ket{\psi(\{x_a(t)\})}$ and $\ket{\varphi(t)}$ form sides of a triangle with the angle between the exact quantum wave function and its TDVP approximation equal to the Fubini-Study distance between those two states. Using the law of cosines,
\begin{equation}
  \gamma(\psi(\{x_a(t)\}),\Psi) \le \begin{cases}
    \frac{\pi}{2} & \text{if } I_t > \sqrt{2}\text{,} \\
    \arccos \left(1 - {I_t^2}/{2}\right) & \text{otherwise.}
  \end{cases}
\end{equation}

Assuming that we initialized the system on the periodic TDVP trajectory, and taking the time $t=T$ to be the period of this trajectory, we  arrive at the following bound on fidelity,
\begin{equation}\label{Eq:bound-finite}
  \sqrt{F_\Psi} \ge 1-\frac{I_T^2}{2},
\end{equation}
where $F_\Psi = | \langle \psi(\{x_a(T)\}) | \Psi(t)\rangle  |^2 =| \langle \Psi |e^{-iHT}| \Psi \rangle  |^2$ and we used the fact that the quantum system is initialized on the TDVP periodic trajectory, $\ket{\Psi(0)}=\ket{\psi(\{x_a(0)\})}=\ket{\psi(\{x_a(T)\})}$.

Until now, we did not discuss the scaling of fidelity and error with the system size. Na\"ively if we take the thermodynamic limit in Eq.~(\ref{Eq:bound-finite}), this bound would appear useless since $I^2_t  \propto L $ increases linearly with the system size and eventually becomes larger than one. However, assuming that the entanglement growth is not too fast (the weak leakage case), one may find an intermediate system size such that $I^2_t\ll1$ and the bound is still effective.  In such a case, assuming the scaling form of the fidelity $F_\Psi  = e^{-f_T L}$, with $f_T\ll1$, we can expand  $\sqrt{F_\Psi}  = 1-f_T L/2$. Plugging this expansion in Eq.~(\ref{Eq:bound-finite}) along with the definition of $\Gamma_T = [\int_0^T dt \gamma(t)]^2  = I^2_t/L$, we obtain, 
\begin{equation}\label{Eq:fsimple}
1-\frac{f_T L}2 \geq 1-\frac{\Gamma_T L}{2}
\end{equation}
 recovering the upper bound on $f_T$ that governs the fidelity decay, see Eq.~(\ref{Eq:fid-bound}) in the main text.

We note that the above argument cannot be regarded as a rigorous proof of the fidelity bound in the many-body case. However, using the limited velocity of entanglement growth from the initial weakly-entangled state $\ket{\psi(\{x_a(T)\})}$, we can limit the system size $L$ needed to be effectively in the thermodynamic limit, thus justifying the expansion used in deriving Eq.~(\ref{Eq:fsimple}). It is an interesting open question if a similar bound can be rigorously proven under some assumptions limiting the entanglement growth, and if a tighter bound could be obtained.  

\begin{figure}[tb]
\begin{center}
\includegraphics[width=0.65\columnwidth]{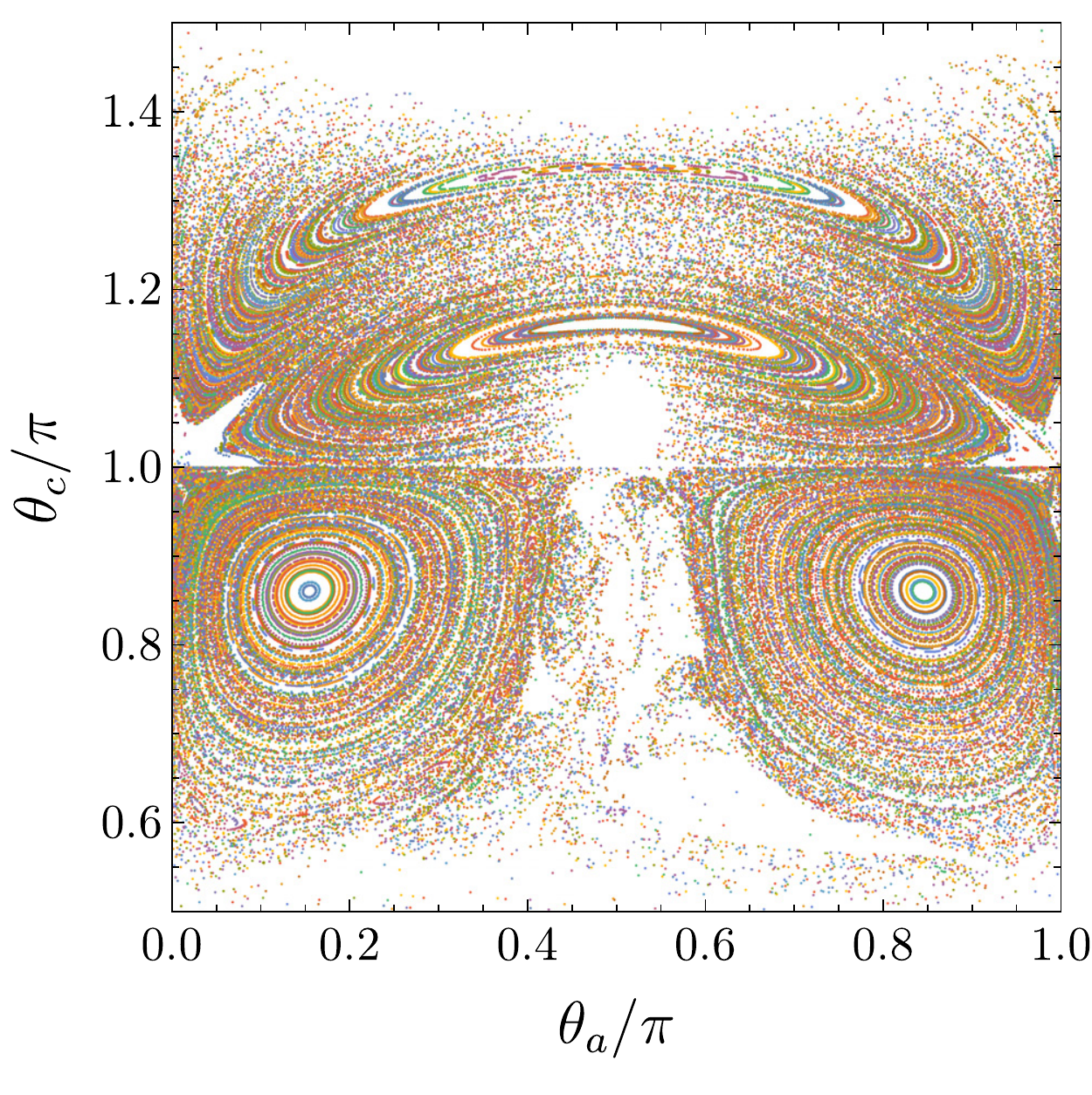}
\caption{ \label{Fig:deformed-PS} 
 Poincar\'e section $(\theta_2=0,\dot\theta_2<0)$ for the deformed PXP model with $\mu_3 =-0.25$ shows an overall increase of the regular islands.}
\end{center}
\end{figure}
 
\begin{figure*}
\begin{center}
\includegraphics[width=1.99\columnwidth]{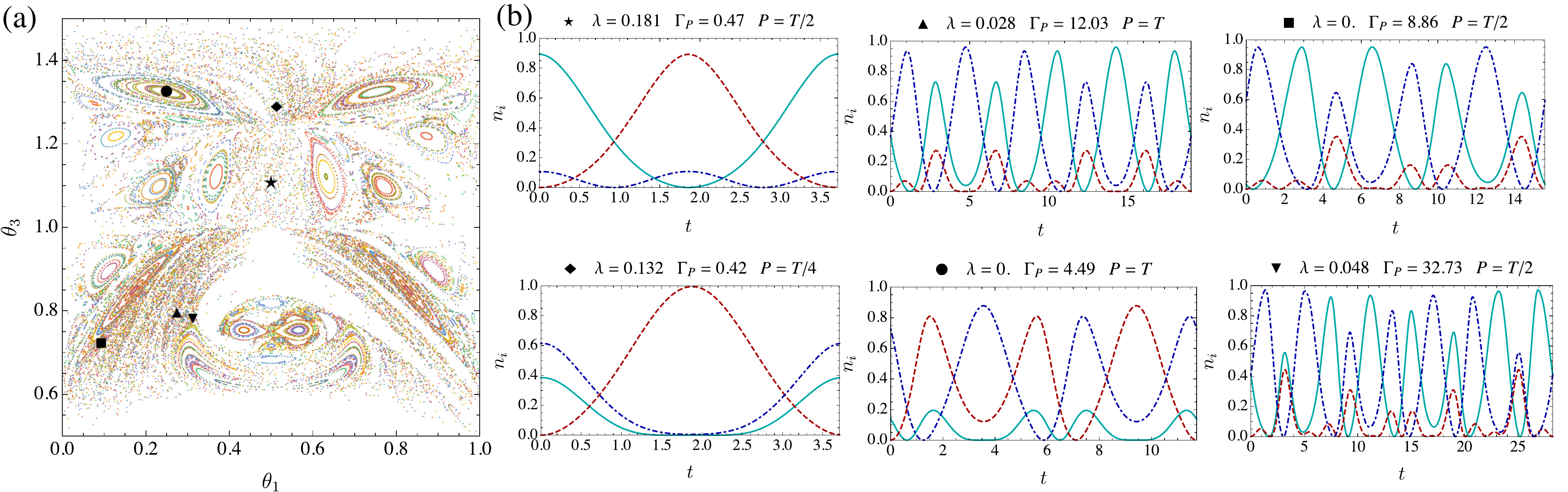}\\
\caption{ \label{Fig:tra-analyze} TDVP analysis of trajectories in the deformed PXP model with $\mu_3 =0.25$. (a) Poincar\'e section $(\theta_2=0,\dot\theta_2<0)$ reveals a smaller size of regular islands in phase space. Black symbols $\star,\blacklozenge,\blacktriangle,\bullet,{\small \blacksquare},\blacktriangledown$
 show the locations of some periodic orbits. For each of the orbits, in panel (b) we show the value of the Floquet exponent, quantum leakage, $\Gamma_P$, and the dynamics of local observables $n_{1,2,3}$. The physical period, $P$, is defined as the time after which the local observables return to their initial values and can be smaller than the period of the orbit for the MPS parameters. 
}
\end{center}
\end{figure*}

\subsection{Trajectories in the deformed PXP model with $K=3$ ansatz \label{App:PXP-trajectories}}
\subsubsection{Calculation of leakage}
In what follows we use the leakage function calculated for the case of the Hamiltonian given by Eqs.~(\ref{Eq:PXP}) and~(\ref{Eq:h3}). To calculate the leakage we use Mathematica to perform the required contractions and substitute them into Eqs.~(\ref{Eq:GramMatfull})-(\ref{Eq:Hamsq}). While in the case of the non-deformed PXP model one can obtain simple analytical expressions for the leakage, for the deformed PXP model these expressions become too lengthy to be presented here~\SOM. 

\subsubsection{Calculation of Floquet exponent}

In order to characterize the periodic orbits observed in the TDVP dynamics we use the integrated leakage that was discussed above. In addition, we study the stability of periodic orbits by calculating the Floquet exponent. The Floquet exponent characterizes the stability of a given periodic orbit, in a manner similar to the Lyapunov exponent. However, despite qualitative similarities, the Floquet exponent quantitatively differs from the Lyapunov exponent~\cite{ChaosBook}. The latter quantity was used to characterize chaotic dynamical flow in the case of large bond dimension TDVP~\cite{Green18}. In contrast, here we are dealing with periodic orbits. In this case, the Floquet exponent allows to quantify the orbit stability in a way that is invariant under all local smooth nonlinear coordinate transformations. Thus, the Floquet exponent is an intrinsic characteristics of the periodic orbit. 

Since we do not use the Lyapunov exponent in this work, we keep the notation $\lambda$ for the Floquet exponent. It is defined as
\begin{equation}\label{Eq:mu-def}
\lambda = \frac{1}{T}\ln |\Lambda_\text{max}|,
\end{equation}
where $\Lambda_\text{max}$ is the largest-norm Floquet multiplier of the orbit in the space of TDVP parameters, $x_a^{(0)}(t)$, and $t\in [0,T]$ where $T$ is the orbit period. Intuitively, $\Lambda_\text{max}$ characterizes how the unit volume, chosen at some initial point on the periodic orbit, transforms after one traversal of the orbit. More specifically, $\Lambda_\text{max}$ is the largest-norm eigenvalue of the orbit Floquet matrix, $[J_T]_{ab} = \partial x_a(T)/\partial x_b(0)$, that quantifies the effect of small perturbation at time $t=0$ after one period of the orbit, $T$. The orbit Floquet matrix is calculated as the time-ordered integral of the instantaneous Jacobian matrix along the periodic orbit~\cite{ChaosBook}, 
\begin{equation}\label{Eq:F-def}
 [J_T]_{ab} = {\cal T}\exp\left[ \int^T_0 dt\, \left.\frac{\partial v_a(x)}{\partial x_b}\right|_{x\to x^{(0)}(t)}\right]
\end{equation}
where we assumed that $v_a(x)$ is the vector of velocities that generates the flow [in other words, the equations of motion are written as $\dot x_a = v_a(x)$]. 

Numerically, the time-ordered integral in Eq.~(\ref{Eq:F-def}) is calculated by solving a system of non-autonomous first-order differential equations. Since we have three variational parameters $\theta_{1,2,3}$, $J_T$ is $3\times 3$ matrix.  As expected, we find that the spectrum of $J_T$ always has an eigenvalue $\Lambda=1$~\cite{ChaosBook}, and the remaining two eigenvalues satisfy $|\Lambda_\text{max}\Lambda_\text{min}|=1$, reflecting the symplectic (volume-preserving) nature of the flow. 

If an orbit is stable and surrounded by KAM torus, we find that all eigenvalues of the Floquet matrix have $|\Lambda|=1$. Thus, the Floquet exponent vanishes, $\lambda=0$, signaling the stability of the orbit to small deformations. In contrast, for unstable periodic orbits we have $|\Lambda_\text{max}|>1$ and $|\Lambda_\text{min}|<1$. This implies that a unit volume after one period is elongated in one direction and compresses in a different direction. 

\subsubsection{Trajectory analysis}

With the expressions for quantum leakage and Floquet exponent at hand, we turn to the analysis of different trajectories in the deformed PXP model. First, we consider the deformation with a large negative value of $\mu_3=-0.25$. Figure~\ref{Fig:deformed-PS} shows that this deformation leads to a drastic increase of the size of regular islands compared to Fig.~\ref{Fig:Z3-PS}(a), which shows the Poincar\'e section for $\mu_3=0$. Nevertheless, despite an increase in size of the regular regions, the fidelity revivals are degraded by the deformation with large negative values of $\mu_3$, see Fig.~\ref{Fig:Z3-chaos}. The leakage follows the same trend, as is apparent from the same figure. However, the orbit remains stable, hence its Floquet exponent $\lambda =0$ throughout the entire range of negative values of $\mu_3$. 

Next, we turn to the case of $\mu_3 = 0.25$. The Poincar\'e section in Fig.~\ref{Fig:tra-analyze}(a) shows that the deformation with $\mu_3>0$ decreases the size of regular regions and makes the TDVP dynamics more chaotic. Different symbols in Fig.~\ref{Fig:tra-analyze}(a) show the location of several periodic orbits with short periods that we were able to find using Newton's search method~\cite{strogatz,ChaosBook} (we show only a subset of short orbits and omit those which are symmetry-related). 

Figure~\ref{Fig:tra-analyze}(b) visualizes the TDVP dynamics of local observables in these different orbits and also shows the integrated leakage, $\Gamma_P$, and the Floquet exponent. In particular, $\star$ symbol shows the location of an orbit related by symmetry to the one studied in the main text. We observe that this orbit is not surrounded by a torus any more, consistent with finite value of $\lambda$. However, there are still many stable orbits present in the Poincar\'e section (e.g., the ones denoted by $\bullet$ and ${\small \blacksquare}$ symbols). These orbits have much stronger leakage, despite the vanishing Floquet exponent $\lambda$. This confirms our conclusion that the Floquet exponent is not obviously associated with quantum leakage. 

On the other hand,  the  $\blacklozenge$  orbit in Fig.~\ref{Fig:tra-analyze}(b) has a slightly smaller leakage compared to the $\star$ orbit. By comparing quantum dynamics we indeed find that these two orbits have similar behavior of the fidelity revivals. However, due to relatively large leakage, the fidelity revivals and oscillations of local observables are damped. Hence, while the orbit $\blacklozenge$ in principle represents another type of oscillations, it remains an open question if the quality of these oscillations can be improved, e.g. if they can be stabilized by a deformation of the Hamiltonian. 

\section{Thermalization in the strong leakage regime}

In this Appendix we present additional results for the cases when the system has no low-leakage trajectories. We start with  additional data for the TFIM. In addition, we consider a strongly deformed PXP model with broken particle-hole symmetry and demonstrate that it still has entanglement dynamics that depends on the initial conditions. 

\subsection{Entanglement dynamics and ETH indicators in TFIM\label{App:TFIM2}}

\begin{figure}[tb]
\begin{center}
\includegraphics[width=0.95\columnwidth]{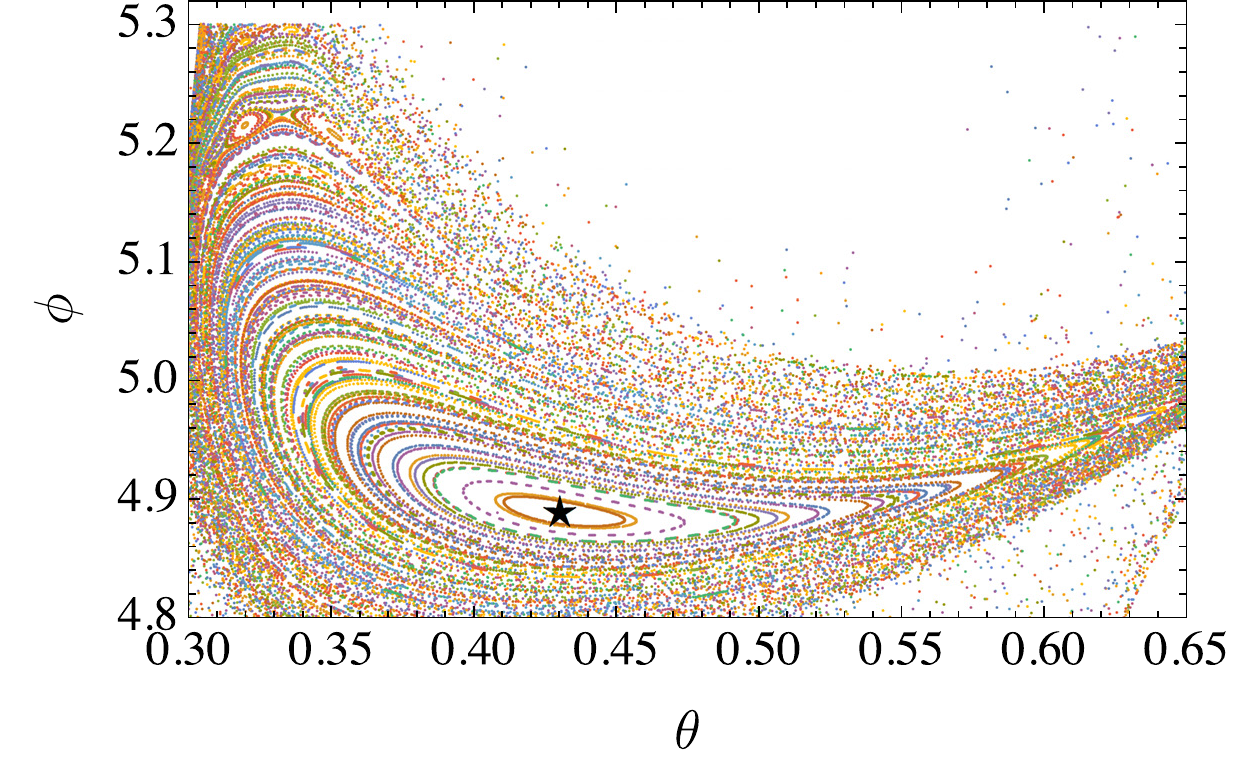}\\
\caption{ \label{Fig:TFIM}
The projection of the Poincar\'e section  onto $(\theta,\phi)$ plane in vicinity of the periodic orbit reveals a regular region of phase space.}
\end{center}
\end{figure}

We begin with illustrating the periodic orbit found in the TDVP dynamics of a TFIM using Newton's search algorithm.  Figure~\ref{Fig:TFIM} illustrates the Poincar\'e section defined by $(\xi=0.9,\dot\xi<0)$. The periodic orbit with period $T= 2.09$ crosses this plane at the point $(\chi,\xi,\phi,\theta ) = (0.2607,0.9,4.888,0.4308)$, shown by   ``$\star$'' symbol in Fig.~\ref{Fig:TFIM}. This is a stable orbit surrounded by a small KAM torus, but nevertheless it does not give rise to many-body fidelity revivals. This may be attributed to large leakage, $\Gamma_T = 0.57$.

\begin{figure}[tb]
\begin{center}
\includegraphics[width=0.999\columnwidth]{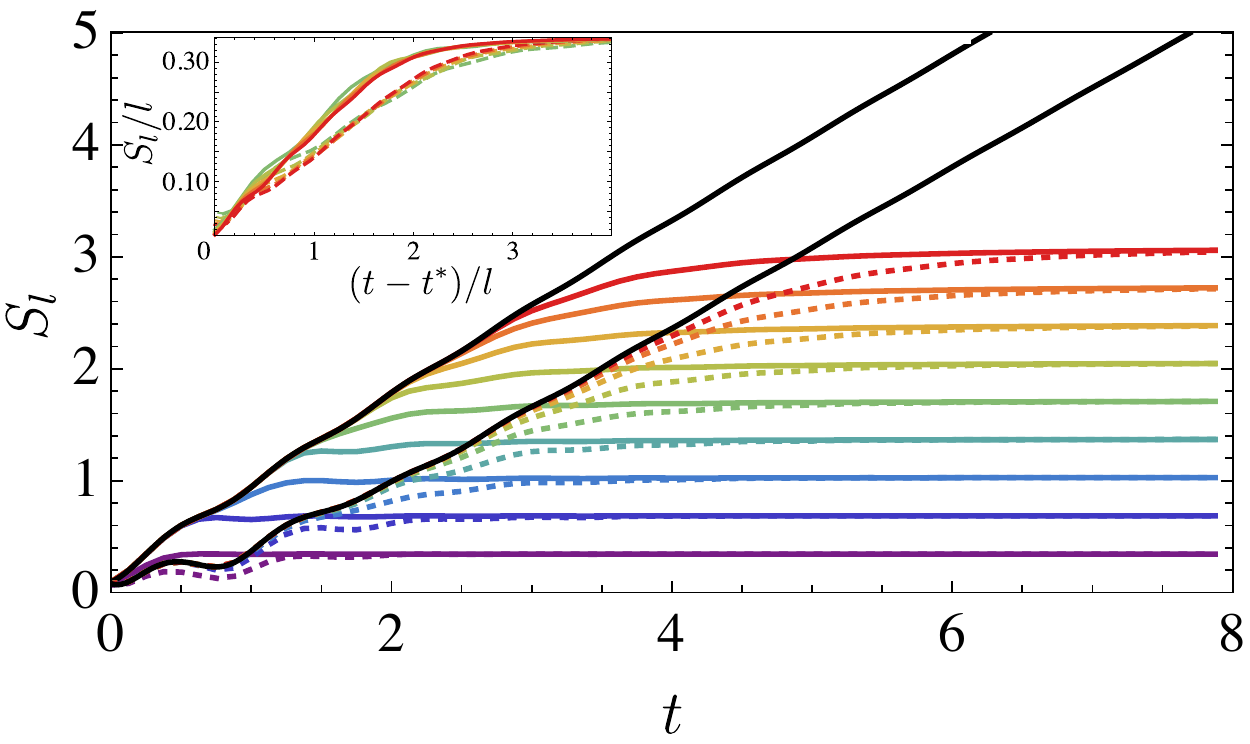}\\
\caption{ \label{Fig:TFIM-sat}
Dynamics of entanglement of subsystems of size $l$ at one boundary, $S_{l} = S(l)/2$, in TFIM for  ``fast'' (solid lines) and ``slow'' (dashed lines) initial states defined in the text. Different colors correspond to different subsystem sizes that range from $l=1$ (magenta) to $l=9$ (red). Black lines correspond to the entanglement entropy of a half-infinite subsystem. The data is obtained with iTEBD with truncation error $\epsilon < 10^{-5}$.The inset shows the same data for $l=5,\ldots, 9$ with rescaled axes, highlighting that the difference in entanglement spreading cannot be explained by the constant time shift and persists for largest systems.}
\end{center}
\end{figure}

In the main text we illustrated the strong influence of initial conditions on  entanglement growth. Here, we compare the initial states with slowest and fastest entanglement dynamics. These states are defined by the MPS in Eq.~(\ref{Eq:TFIM-ansatz}), with parameters $(\chi,\xi,\phi,\theta)_\text{slow} = (0.1,0.9,1.17,3.86)$ and $(\chi,\xi,\phi,\theta)_\text{fast} = (-0.32, 0.9, 1.6, 4.28)$. These states can also be  identified by local expectation values, $(\sigma^{x},\sigma^{y},\sigma^{z})_\text{slow}=(0.38,0.85,0.072)$ and $(\sigma^{x},\sigma^{y},\sigma^{z})_\text{fast}=(0.27,0.74,-0.49)$. Figure~\ref{Fig:TFIM-sat} shows the dynamics of entanglement of small subsystems for slow and fast initial states. Both states reach  identical values of the saturated entanglement entropy for all subsystems. This is expected, since both initial states have the same energy density per site. As energy is the only conserved quantity, these states are expected to have the same entanglement saturation value. However, the time when the entanglement saturates is different, proving that these two initial states have different entanglement velocity. 

\begin{figure}[tb]
\begin{center}
\includegraphics[width=0.95\columnwidth]{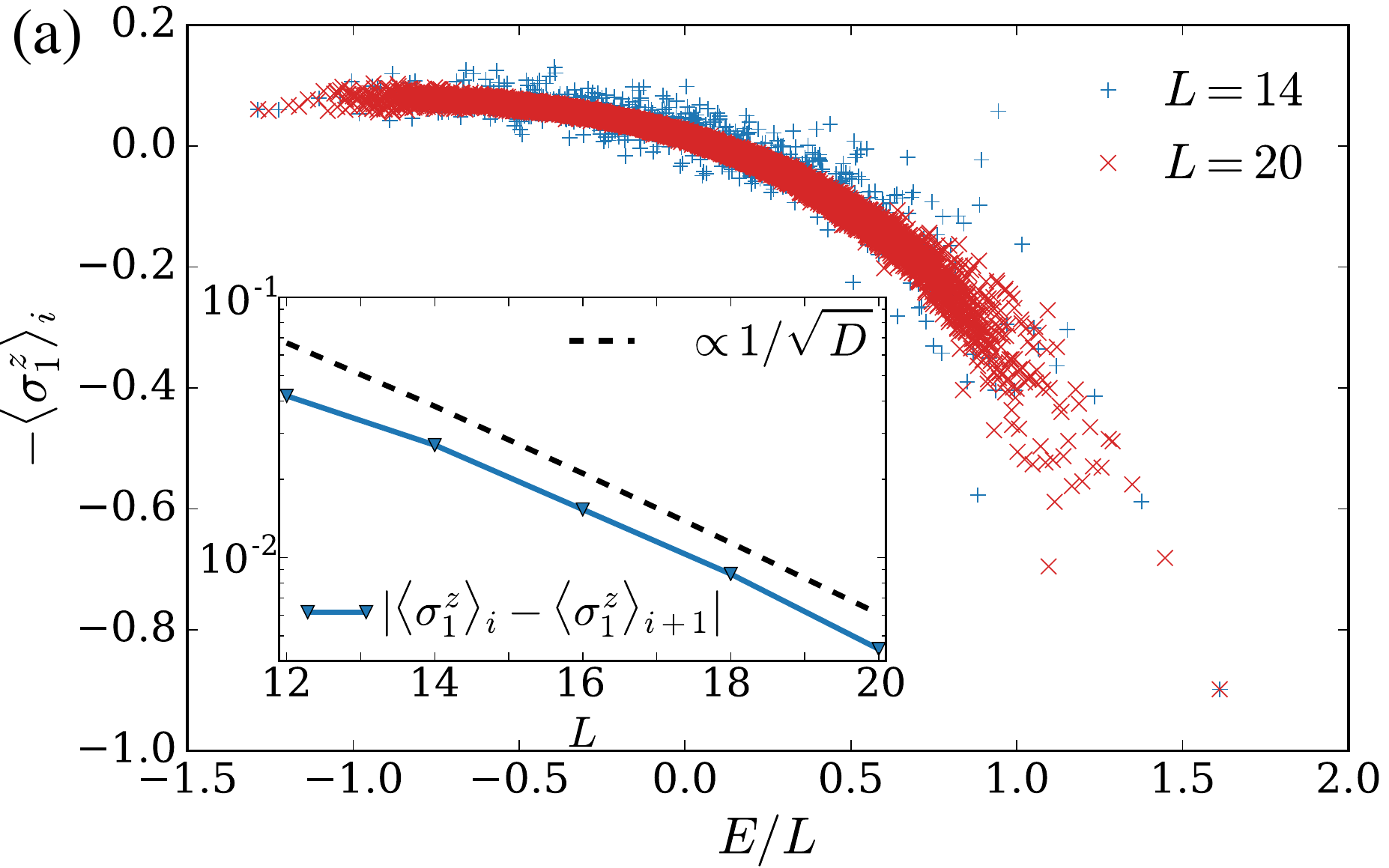}
\includegraphics[width=0.95\columnwidth]{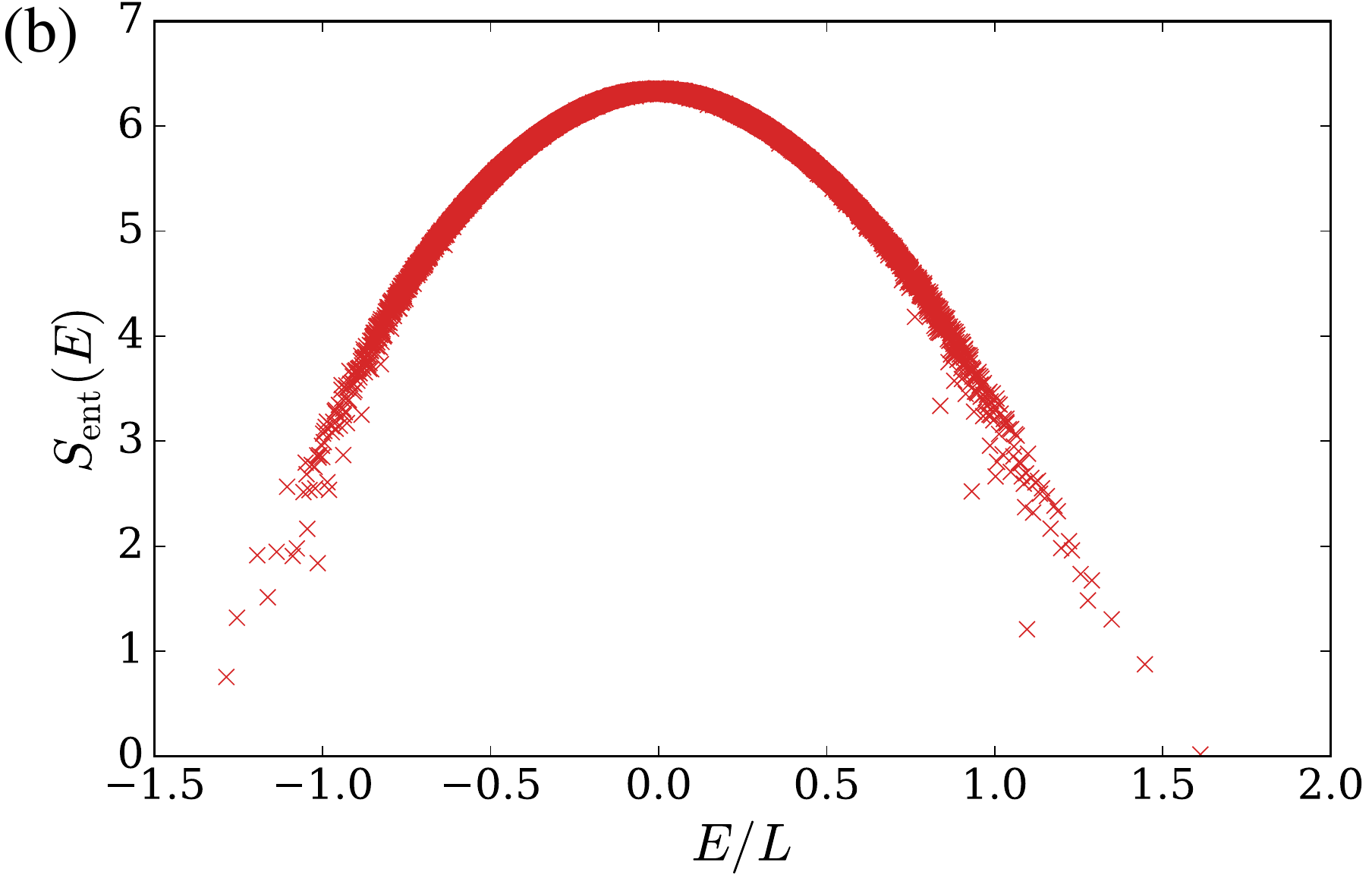}
\caption{ \label{Fig:TFIM-ETH}
(a)  {Comparing the expectation values of magnetization per spin between different system sizes suggest the applicability of ETH. Inset confirms that the fluctuations of magnetization between adjacent eigenstates in the energy window $0.19L+[-1,1]$ are suppressed, in accordance with the ETH prediction, as $1/\sqrt{D}$, where $D$ is the Hilbert space dimension.}
(b) Bipartite entanglement entropy of the eigenstates for $L=20$ shows no outliers near the middle of the spectrum. All data is obtained for zero-momentum inversion-symmetric sector of TFIM with PBC using exact diagonalization.
}
\end{center}
\end{figure}

 {Finally, Fig.~\ref{Fig:TFIM-ETH} shows a test of ETH. For the TFIM for the considered values of parameters, i.e., $(J_z, h_z,h_x) = (1,0.4,1)$. Fig.~\ref{Fig:TFIM-ETH}(a) illustrates the expectation values of local observables and confirms that average differences of local magnetization between adjacent eigenstates at energy density $E/L=0.19$ decay with systems size according to the ETH prediction. In Fig.~\ref{Fig:TFIM-ETH}(b) we observe that the entanglement of eigenstates essentially has no outliers near the middle of the spectrum.
}

\subsection{Entanglement dynamics and ETH indicators in the deformed PXP model  \label{App:PXP-deform}}

\begin{figure}[tb]
\begin{center}
\includegraphics[width=0.95\columnwidth]{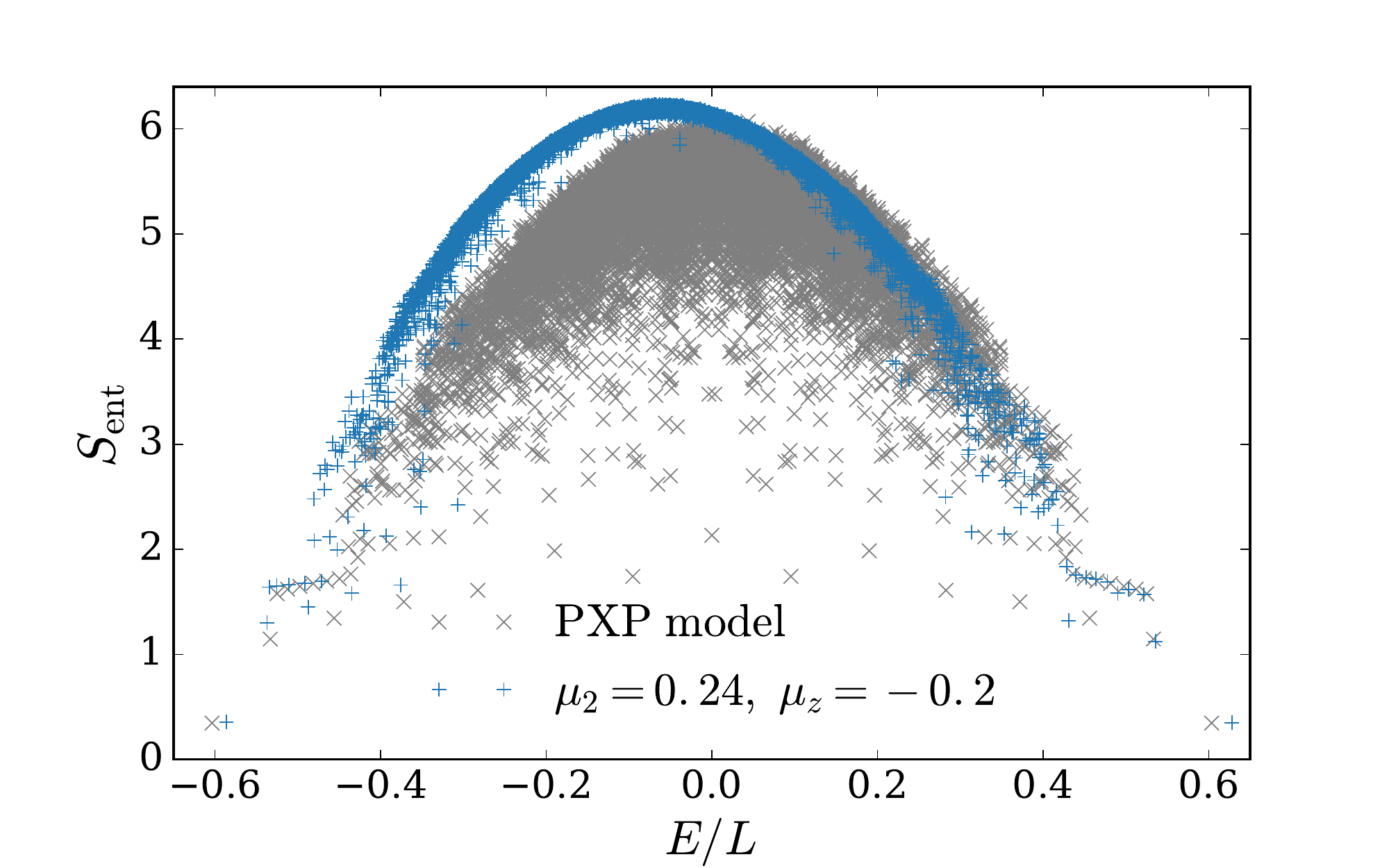}\\
\caption{ \label{Fig:PXPETH}
The deformation $\mu_2=0.24$ and $\mu_z = 0.4$ effectively destroys the eigenstates with low bipartite entanglement entropy in the pure PXP model. The data is obtained for $L=28$ chain with PBC in zero momentum inversion-symmetric sector.}
\end{center}
\end{figure}

We consider the following deformation of the PXP model, $\delta H  = H_{\mu_z}+ H_{\mu_2}$ where
\begin{equation}\label{Eq:PXP2}
H_{\mu_2} =  \mu_2\sum_i \left(P_{i-2}\sigma^+_{i-1}P_i\sigma^+_{i+1}P_{i+2} + \text{h.c.}\right),
\end{equation}
and values of the couplings are fixed to $\mu_2=0.24$ and $\mu_z = 0.4$. This deformation is chosen in a such way that both terms in $\delta H$ break the particle-hole symmetry of the model. In addition, the correlated spin flip term is expected to enhance thermalization. 

Indeed, the entanglement entropy of many-body eigenstates of the Hamiltonian $H_{PXP}+\delta H$ shown in Fig.~\ref{Fig:PXPETH} shows that there are no ``outliers'', i.e., states with anomalously low entanglement. Likewise, the level statistics indicator $r= {\min(\delta_i,\delta_{i+1})}/{\max(\delta_i,\delta_{i+1})}$ yields $\corr{r} = 0.528$, a value  that is close to the GOE prediction $0.5307$~\cite{Atas}~(averaging is done over the middle 1/3 of the full many-body spectrum). 

\begin{figure}[b]
\begin{center}
\includegraphics[width=0.95\columnwidth]{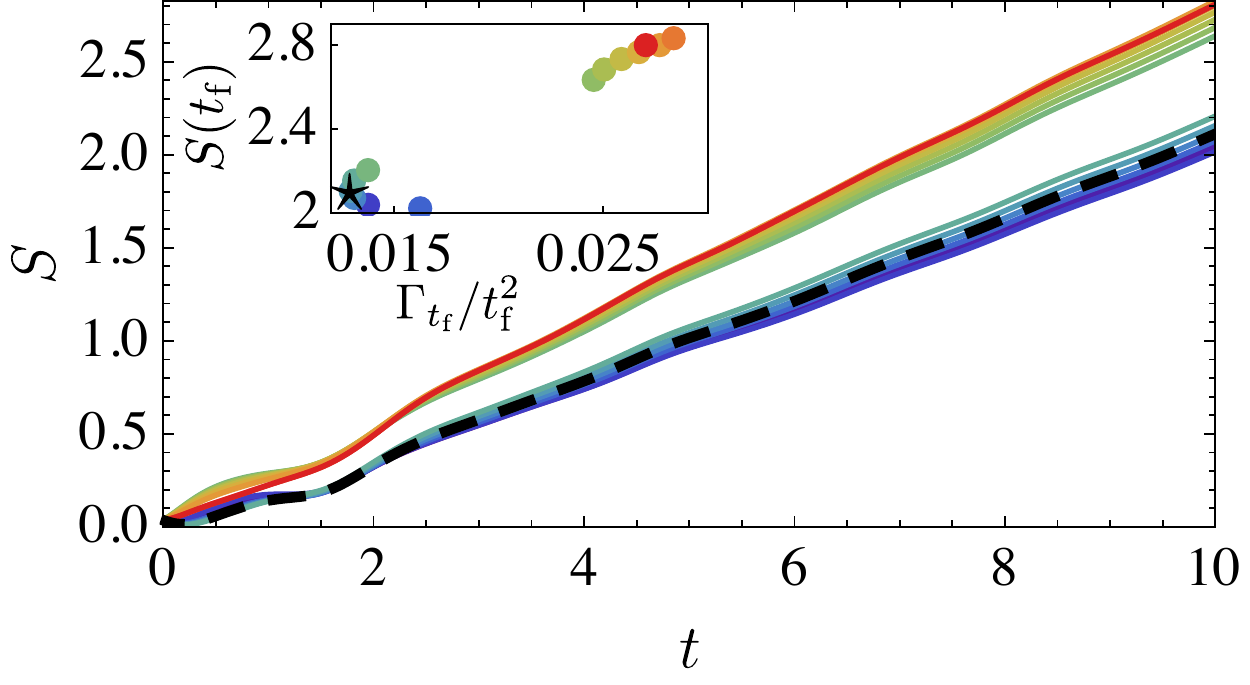}\\
\caption{ \label{Fig:thermo-PXP}
Entanglement growth in the deformed PXP model, Eq.~(\ref{Eq:PXP2}), strongly depends on initial conditions. All initial conditions are chosen to have the same energy density $\braket{H}/L = 0$ and the same entanglement entropy. The light blue line corresponds to the periodic trajectory. The inset shows the correlation between the value of entanglement at late time and quantum leakage $\Gamma_{t_f}/t_{f}^2$, $t_{f} = 10$.
}
\end{center}
\end{figure}

The TDVP equations of motion for $K=2$ MPS ansatz in presence of this deformation can be obtained using the framework laid out in Appendix~\ref{App:eom}. Using the expectation value of $H_{\mu_2}$ that is given by 
\begin{equation}
\corr{H_{\mu_2}} =-\frac{L}{2}\left(\frac{\cos^2\theta_{2}\cos(2\phi_{2})\sin^2(2\theta_{1})}{2+2\cos^2\theta_{2}\tan^2\theta_{1}}+1\leftrightarrow 2\right),
\end{equation}
we obtain additional terms that have to be added to the equations of motion~(\ref{Eq:Z2-dyn}) 
\begin{equation}
\begin{split}
&\delta \dot \theta_1 =  \mu_2\big(2 \cos^3\theta_{1}\sin\theta_{1}\sin^2\theta_{2}\sin(2\phi_{1})+\\
&\cos^2\theta_{2}\sin(2\theta_{1})\sin(2\phi_{2})\big),\\
&\delta \dot \phi_1 = \mu_2\Big((-3+\cos(2\theta_{1}))\cos^2\theta_{2}\sin^2\theta_{1}\cos(2\phi_{2})-\\
&\frac{1}{2}\cos^2\theta_{1}\cos(2\phi_{1})\big((\cos(2\theta_{1})-5)\cos(2\theta_{2})+2\sin^2\theta_{1}\big)\Big).
\end{split}
\end{equation}
The EOMS for $\delta \dot \theta_2 $, $\delta \dot \phi_2 $ are obtained by swapping the subscripts $1\leftrightarrow 2$.

Similarly to the case of TFIM discussed in the main text, we study the dependence of entanglement dynamics on the initial condition within the MPS manifold. Initially, we find a periodic trajectory at zero energy density. The trajectory is found by starting from the $\mathbb{Z}_{2}$ trajectory of the PXP model in a similar fashion to the trajectory with the chemical potential discussed in the main text. In this case, we start from the periodic trajectory at $\mu_{z} = 0.4$, and slowly ramp up $\mu_{2}$ at fixed energy density, $\braket{H}/L = 0$. In practice, we can choose the initial state to be any phase space point of the periodic trajectory. We have observed that even though the entropy of states in this periodic trajectory fluctuates, it does not considerably affect the long time entropy growth in the exact quench dynamics. Therefore, we fix the initial point to be $(\theta'_{1},\phi'_{1},\theta'_{2},\phi'_{2})= (3.926, 0.289, 2.987, 0.121)$. This state has bipartite entanglement entropy $S\approx 0.036$. In the chosen MPS ansatz the bipartite entanglement entropy $S(\theta_{1},\theta_{2})$ is independent of the phases. Thus, we can easily pick different initial states with exactly the same entropy by varying $\phi_{2}$ while $\phi_{1}$ is fixed by the energy density constraint. The initial states of Fig.~\ref{Fig:thermo-PXP} are generated by splitting the domain $[\phi'_{2}-\pi/4,\phi'_{2}+\pi/4]$ to $200$ points, and then discarding the points for which the energy density constraint cannot be satisfied for any value of $\phi_{1}$. The number of remaining states are $15$, including the periodic trajectory. 

In Fig.~\ref{Fig:thermo-PXP} we observe that the entropy growth of a state is closely related to the leakage of the state out of the MPS manifold. Compared to the case of TFIM studied in the main text, the PXP model features stronger correlations between leakage and entanglement with correlation coefficient being 0.95 (compared to 0.53 for TFIM). In this case there are two sets of states which have different velocities of entropy growth, and all the states in the ``slow" set have considerably smaller leakage than any of the states in the ``fast" set. The periodic trajectory belongs to the set of slow states but it is not the slowest one.

\end{document}